\documentclass[a4paper,11pt]{article}
\usepackage[dvipsnames]{xcolor}
\usepackage[utf8]{inputenc}
\usepackage[bbgreekl]{mathbbol}
\usepackage{geometry}
\usepackage{multicol}
\usepackage{float} 

\usepackage{extarrows}
\usepackage{xcolor}
\usepackage{amsmath, array, amssymb, amsfonts,amsthm}
\usepackage{upgreek}
\usepackage{xfrac}
\usepackage[inline]{enumitem}
\setlist{nolistsep}
\usepackage[french, italian, english]{babel}
\usepackage[all]{xy}
\usepackage{txfonts}  
\usepackage{sectsty}
\usepackage{booktabs}
\usepackage{caption}
\usepackage{dsfont}
\usepackage{mathtools}
\usepackage{slashed}
\usepackage[makeroom]{cancel}
\usepackage[hidelinks]{hyperref}
\usepackage{textcomp}
\usepackage{multicol}
\usepackage[title]{appendix}
\usepackage[square,numbers,sort,compress,comma,merge]{natbib}
\usepackage{hyperref}
\hypersetup{colorlinks=true, urlcolor=blue, citecolor=blue, linktoc=page}

\usepackage{tikz}
\usepackage{tikz-cd}
\usetikzlibrary{cd}
\tikzcdset{
arrow style=tikz,
diagrams={>={Straight Barb[scale=0.8]}}
}
\usetikzlibrary{matrix,arrows,decorations.pathmorphing}

\DeclareMathAlphabet{\mathpzc}{OT1}{pzc}{m}{it}

\usepackage{fancybox}

\usepackage{mathabx}
\usepackage{mathrsfs}
\usepackage{scrextend}

\makeatletter
\renewcommand*\env@matrix[1][\arraystretch]{%
  \edef\arraystretch{#1}%
  \hskip -\arraycolsep
  \let\@ifnextchar\new@ifnextchar
  \array{*\c@MaxMatrixCols c}}
\makeatother

\newcommand{\defeq}{\vcentcolon=}
\newcommand{\rdefeq}{=\vcentcolon}

\newcommand\M{\mathcal{M}}
\newcommand\N{\mathcal{N}}

\newcommand\RR{\mathbb{R}}
\newcommand\CC{\mathbb{C}}

\newcommand\C{\mathcal{C}}

\newcommand\id{\textit{id}}

\newcommand\G{\mathcal{G}}
\renewcommand\H{\mathcal{H}}

\renewcommand\S{\mathcal{S}}

\newcommand\SU{\mathcal{SU}}
\newcommand\U{\mathcal{U}}
\newcommand\SO{\mathcal{SO}}
\newcommand\SL{\mathcal{SL}}

\renewcommand\O{\mathcal{O}}
\newcommand\GL{\mathcal{GL}}

\newcommand\W{\mathcal{W}}
\newcommand\vphi{\varphi}

\renewcommand\u{\text{\bf u}}

\renewcommand\epsilon{\varepsilon}

\newcommand\rarrow{\rightarrow}

\newcommand\LieG{\mathfrak{g}}
\newcommand\LieH{\mathfrak{h}}

\newcommand\so{\mathfrak{so}}

\renewcommand\b{\bar }
\newcommand\w{\wedge}
\renewcommand\d{\partial}
\newcommand\s{\sigma}
\newcommand\bs{\boldsymbol}

\renewcommand\-{^{-1}}
\newcommand\Ad{\text{Ad}}
\newcommand\ad{\text{ad}}
\renewcommand\id{\text{id}}

\makeatletter

\newcommand{\Rmnum}[1]{\expandafter\@slowromancap\romannumeral #1@}
\makeatother

\makeatletter
\newcommand{\leqnomode}{\tagsleft@true\let\veqno\@@leqno}
\newcommand{\reqnomode}{\tagsleft@false\let\veqno\@@eqno}
\makeatother

\DeclareMathOperator{\Diff}{Diff}
\DeclareMathOperator{\Aut}{Aut}

\DeclareMathOperator{\Tr}{Tr}

\DeclareMathOperator{\im}{Im}

\theoremstyle{definition}


\begin{document}


\title{What Price Fiber Bundle Substantivalism? \\ On How to Avoid Holes in Fibers}

\author{P. \textsc{Berghofer} $\,{}^{a,\,*}$ \and J. \textsc{François} $\,{}^{a,\, b,\, c,\,\dagger}$ \and L. \textsc{Ravera} $\,{}^{d,\,e,\,f,\,\star}$}

\date{}

\maketitle
\begin{center}
\vskip -0.6cm
\noindent
${}^a$ Department of Philosophy, University of Graz -- Uni Graz. \\
Heinrichstraße 26/5, 8010 Graz, Austria.\\[2mm]
 
${}^b$ Department of Mathematics \& Statistics, Masaryk University -- MUNI.\\
Kotlářská 267/2, Veveří, Brno, Czech Republic.\\[2mm] 
 
${}^c$ Department of Physics, Mons University -- UMONS.\\
20 Place du Parc, 7000 Mons, Belgium.
\\[2mm]

${}^d$ DISAT, Politecnico di Torino -- PoliTo. \\
Corso Duca degli Abruzzi 24, 10129 Torino, Italy. \\[2mm]

${}^e$ Istituto Nazionale di Fisica Nucleare, Section of Torino -- INFN. \\
Via P. Giuria 1, 10125 Torino, Italy. \\[2mm]

${}^f$ \emph{Grupo de Investigación en Física Teórica} -- GIFT. \\
Universidad Cat\'{o}lica De La Sant\'{i}sima Concepci\'{o}n, Concepción, Chile. \\[2mm]

\vspace{1mm}

${}^*$ {\small{philipp.berghofer@uni-graz.at}} \quad \quad
${}^\dagger$ {\small{jordan.francois@uni-graz.at}} \quad \quad ${}^\star$ {\small{lucrezia.ravera@polito.it}}
\end{center}


\vspace{-3mm}

\begin{abstract}

On a mathematically foundational
level, our most successful physical theories (gauge field theories and general-relativistic theories) are formulated in a framework based on the differential geometry of connections on principal bundles. After reviewing the essentials of this framework, we articulate the generalized hole and point-coincidence arguments, examining how they weight on a substantivalist position towards bundle spaces. This question, then, is considered in light of the Dressing Field Method, which allows a manifestly invariant reformulation of gauge field theories and general-relativistic theories, making their conceptual structure more transparent: it formally implements the point-coincidence argument and thus allows to define (dressed) fields and (dressed) bundle spaces immune to hole-type arguments. 

\end{abstract}


\vspace{-3mm}

\tableofcontents

\section{Introduction}

One of the central aims of philosophy of science, and philosophy of physics especially, is to determine how scientific theories relate to reality. A basic goal, in particular, is to identify the fundamental ontology suggested by our best physical theories. Sometimes, this task is relatively straightforward---or at least appears to be---as in the case of classical, or even special relativistic, mechanics. At other times, it is highly non-trivial and continues to puzzle even the most careful thinkers for decades. The prime example is quantum mechanics. For instance, whether the wave function is ontologically real, and what exactly it represents (if anything), remains one of the most debated questions in philosophy of physics.  In between these extremes, we have the time-honored subject of the nature of spacetime in general-relativistic (GR) physics, and in the past 25 years a growing interest in the status and meaning of gauge symmetries, which are at the heart of the Standard Model (SM) of particle physics.

Both these topics are actually deeply related.  
The SM, as a gauge field theory (GFT), is built on the principle of \emph{invariance} under \emph{gauge transformations}, i.e. the idea that the Lagrangian must be invariant under local ``internal" transformations of its field variables. 
This is the \emph{Gauge Principle}.
GR is built on the principle of \emph{covariance} under \emph{diffeomorphisms}, i.e. the idea that the Lagrangian must be covariant under local ``space-time" transformations of its field variables. 
This is the \emph{General Covariance Principle}.
Together, GFT and GR physics constitute the framework of general-relativistic gauge field theory (gRGFT), 
based on principles of local symmetries. 

Regarding the philosophical endeavor mentioned in the opening paragraph, these symmetries pose a significant challenge. 
They complicate the task of identifying which parts of the formalism refer \emph{directly} to basic physical entities.
Indeed, it is  well-established  in physics that physical magnitudes (and observables) ought to be invariants under local transformations---i.e. the latter are \emph{symmetries} of the former.
However, in GFT, the SM in particular, none of the fundamental field variables are gauge-invariant: 
if they refer to physical fields, physical degrees of freedom (d.o.f.) of distinct types, they do so in a non-trivially redundant way.
This raises the question of 
what the meaning of this mathematical over-abundance is, and whether it is possible to describe the relevant physics without it (see, e.g., \cite{Berghofer-et-al2023,Healey2007,Earman2004}). 

Idem for GR, with the added complication that its local transformations, diffeomorphisms, also act on the mathematical object taken to refer to spacetime: the manifold on which the field variables ``live". 
This is the starting point of the literature on the \emph{hole argument} in GR, and the way it gives a new spin on the traditional philosophical dispute between substantivalist and relationalist views of space(time). 
What should be clear from GR is that the ontological status of the fields and that of spacetime are tightly interconnected: this is the fundamental insight of the \emph{point-coincidence argument} (see \cite{Stachel1986} and \cite{Giovanelli2021}).

What is less well appreciated is that the same holds true in GFT: the ontological status of the fields is better understood when put in relation with that of the space in which they ``live" and on which gauge transformations act as well: fiber~bundle spaces. 
Despite being as omnipresent (at least tacitly) in GFT as standard manifolds are in GR, the question of how much of a realist one can be toward fiber bundles has received little attention. 
Even scarcer is the literature that articulates or discusses generalizations of the hole and point-coincidence arguments and examines their implications for a realist view of bundles.
\footnote{Exceptions that treat this topic are \cite{Lyre1999}, \cite{Healey2001}, \cite{Stachel2014}, and \cite{JTF-Ravera2024c}. }

This is the issue we aim to address in a systematic and unified way in this paper. 
We start by reviewing in section~\ref{The hole argument} the hole argument and its implications in GR physics.
In section~\ref{Fiber bundles and basics of gauge theory}, we provide a dense presentation of the mathematical framework underpinning gRGFT: the differential geometry of connections on principal bundles. 
This sets the stage for us to articulate precisely the generalized hole and point-coincidence arguments in section~\ref{The generalized hole argument}, and show how they weight in on the substantivalist \emph{vs} relationalist views of principal bundles. 
As we shall see, the conceptual picture they invite does not exactly follow the dividing lines of the \emph{traditional} debate between relationalism and (sophisticated) substantivalism. This 
is due to how deeply  conceptually entwined relata and relations are. 
Finally, in section~\ref{Dressing Field Method: Dressed fields and spaces} we show that the Dressing Field Method (DFM) allows an invariant reformulation of gRGFTs  that makes their conceptual structure more transparent, and their philosophical analysis much easier: 
In particular, it allows to define \emph{dressed} principal bundles and \emph{dressed} fields which are, by construction, immune to hole-type arguments. 
In section \ref{Conclusion}, we conclude with a few remarks on the interpretive resources of the view presented here.

\section{The hole argument in general-relativistic theory}
\label{The hole argument}

The traditional opposition between the substantivalist and relationalist views on space(-time), famously a topic of the Leibniz-Clarke debate, hinges upon the question as to whether space(-time) has an existence autonomous from the physical objects it ``contains":
Substantivalists (with Newton and Clarke) affirm the proposition, relationalists (with Leibniz) deny it.\footnote{Of course, the word ``contains" is already loaded and tacitly skews towards an a priori substantivalist view.} 
It should be remarked  that the debate is  not necessarily one opposing realist \emph{vs} anti-realist views on space(-time), the substantivalist being the realist and the relationalist being the anti-realist. 
The relationalist can be as much a realist as the substantivalist,  affirming the existence of space(-time) simply not at a primary ontological level, but rather at a secondary, emergent or effective one: it \emph{supervenes} on  ontologically primary physical objects.\footnote{The space(-time) relations  among ontologically primary physical objects being analogous to that of, say, the temperature of a gas or the fluidity of a liquid; an idea resonant with current physics' speculations about an ``emergent" spacetime. 
}

This debate must be reconsidered in light of the advent of General Relativity (GR), and general-relativistic physics more broadly, 
which not only imposes the notion of spacetime---with a dynamics coupled to that of matter-energy---but also forces us to confront the fact that this notion is no longer simple and primitive, but a layered one:
In the best practice of physics, a general-relativistic spacetime is defined as a Lorentzian manifold $(M, g)$, where $M$ is a smooth differential manifold endowed with $g$, a Lorentzian metric. 
The bare manifold $M$ may a priori be understood to represent the totality of ``spatiotemporal events" in their topological and differential relations only.\footnote{The ``bare" manifold is itself not a simple notion. We have first the set-theoretic level, the collection of points as an unstructured set. 
Then we have the topological one: endowing the set with a topology (setting contiguity relations among points and subsets of points) resulting in a topological manifold. 
Finally, endowing the latter with a smooth structure (establishing the possibility of differential calculus) results in a differential manifold. 
The next level would be endowing it with a connection, to get a connected manifold (where geodesics, curvature and torsion may be defined), 
and/or endowing it with a metric, to get a metric manifold. 
Then one may ask compatibility between the connection and metric structures, or remark that a metric uniquely induces a connection, as is often the case in standard GR textbooks.} 
To define their spatiotemporal relations, in particular the causal structure, $g$ is needed.
This leads to both an update and a ``lifting of degeneracy" of the traditional debate, 
which now comes in two related variants:
the substantivalist \emph{vs} relationalist views on \emph{spacetime}, 
and the substantivalist \emph{vs} relationalist views on the \emph{manifold of events}.\footnote{
Famously, \cite{EarmanNorton} endorsed the view that $M$ represents spacetime.
Yet, when physicists are careful and self-aware, they use the word ``spacetime", correctly, to refer to $(M, g)$. 
Only when they are loose with language---as is often the case in the specialized literature---do they tend to (mis)use the term to refer to $M$.
We might be well advised to follow physics best practice, and use a terminology that reflects it. 
We shall then care to distinguish the ``manifold of events" and ``spacetime" variants of the substantival \emph{vs} relationalist debate.}

It should be no surprise that both variants are crucially informed by Einstein's dual
``hole argument" (``\emph{Lochbetrachtung}") and ``point-coincidence argument", as they were key to his timely understanding of the meaning of diffeomorphism covariance leading to the final completion of GR, and to his definite views on spacetime. 
It is unfortunate that subsequently most physicists misunderstood these as philosophical meanderings distracting Einstein from the straight technical path, when they did not just ignore or outright forget them.   
Their deep significance was rediscovered and brought to a wider attention by Stachel in 1980 (published in \cite{Stachel1989}), and the arguments were given a modernized form in \cite{EarmanNorton}.
The literature on the subject is now considerable, and even involves physicists engaged in the open problem of the quantum nature of gravity/spacetime.
Let us give a brief description of the issue, 
 drawing from the account of \cite{JTF-Ravera2024c}.
\medskip

General-relativistic physics describes the coupled dynamics of matter-energy and gravity, both fundamentally understood as \emph{fields}. 
Their mathematical description starts with defining a smooth manifold $M$, representing the  totality of spatio-temporal events. 
It is then endowed with a Lorentzian metric field $g$ representing the gravitational field: the couple $(M, g)$, a Lorentzian manifold, is taken to represent a relativistic spacetime.
Other fields $\upvarphi$ are then defined on $M$, representing matter and interaction fields.
The latter may be thought of as the ``content" of spacetime, or all fields $\upphi=\{g, \upvarphi\}$ may be seen as the ``content" of the manifold of events $M$.
These fields satisfy (coupled) field equations $E(\upphi)=E(g, \upvarphi)=0$ on $M$ or on a finite region $U\subset M$.  
Usually they are derived from an action functional $S(\upphi)=\int_U L(\upphi)$, where $L(\upphi)\in \Omega^m(M)$, with $m=$ dim$(M)$, is a Lagrangian  differential top-form on $M$. 

The mathematical object $M$ has a natural group of automorphisms: the group of diffeomorphisms $\Diff(M)$.
It~acts by pullback on fields on $M$, so that $\upphi \mapsto \upphi^\psi\defeq \psi^* \upphi$ for any $\psi\in \Diff(M)$. 
Any field $\upphi$ thus has a $\Diff(M)$-orbit $\O_\upphi$, and the space of all possible fields $\Phi=\{ \upphi\}$ is partitioned into distinct orbits. 
The space of $\Diff(M)$-orbits is the quotient space $\M\defeq \Phi/\Diff(M)$, called a \emph{moduli space}, and
there is a natural projection $\uppi:\Phi \rarrow \M$, $\upphi \mapsto \uppi(\upphi)\rdefeq [\upphi]$, s.t. $[\phi]=[\phi^\psi]$, i.e. there is a 1:1 correspondence $\O_\upphi \leftrightarrow [\phi]$. 
\medskip

A defining feature of GR physics is the requirement of covariance of (the Lagrangian and thus of) the field equations under $\Diff(M)$:
The~field equations then satisfy $E(\upphi)^\psi\defeq E(\psi^*\upphi)=\psi^*E(\upphi ) =0$. 
The space of solutions $\S\defeq\{\, \upphi \in \Phi \ \text{ s.t. }\, E(\upphi)=0\,\}$ is therefore also partitioned into distinct $\Diff(M)$-orbits $\O_\upphi$, s.t. $\uppi: \S \rarrow \M_\S$, $\upphi\mapsto [\upphi]$, with $\M_\S$ the moduli space of solutions. 
The~immediate dual consequence is that the theory cannot distinguish solutions in the same  $\Diff(M)$-orbit $\O_\upphi$, and is blind to the points and regions of $M$. 

An a priori realism toward the formalism, i.e. a pre-critical ontological commitment to mathematical objects---which is what e.g.  manifold $M$ substantivalism is predicated upon---would thus imply that the theory underdeterminates its ontology.
It amounts to accepting a metaphysical multiplicity of fields $\upphi\in \O_\upphi$ that are physically indistinguishable---something that is disreputable in the natural sciences, and in physics especially, which tends to abide by a form of Ockham's principle of parsimony.

The hole argument is meant to further highlight how this position appears  to lead also  to an ill-defined Cauchy problem, and can be thus formulated: 
{Let us define the (normal) subgroup $\Diff_c(M) \subset \Diff(M)$ of diffeomorphisms with compact support, i.e. $\psi \in \Diff_c(M)$  is s.t. $\psi \neq \id_M$ on a domain $D\subset U \subset M$, and $\psi=\id_M$ on $M/D$---such $\psi$ are sometimes called ``hole diffeomorphisms" or ``hole transformations".
One considers $\upphi, \upphi^\psi \in \O_\upphi\subset \Phi$ s.t. $\psi \in \Diff_c(M)$, so that $\upphi=\upphi^\psi$ on $M/D$, but $\upphi\neq\upphi^\psi$ on $D$.}
Then, given boundary conditions for the fields $\upphi$ at $\d U$, the field equations do not uniquely determine a solution in $U$ since $\upphi, \upphi^\psi \in \O_\upphi\subset \S$. 
This in particular implies a failure of determinism if instead one starts from Cauchy data and $D$ is in the future of the Cauchy surface $\Sigma \subset U \subset M$.
Many physicists and philosophers regard this form of indeterminism as unacceptable---specifically because it is not mandated by the formalism itself but arises from a substantivalist interpretation of it. 
\medskip

{A way out of these predicaments, 
that will be essential to this paper,
is Einstein's \emph{point-coincidence argument}---Norton} \cite{Norton1987} credits Stachel for the name---which consists in the observation that the physical content of the theory (in particular its possible verifications) is exhausted by the pointwise {\emph{coincidental}} values of fields---{by which we mean e.g. the relation ``[value $g(x)$ of $g$ at $x$] $\leftrightarrow$ [value $\upvarphi(x)$ of $\upvarphi$ at $x$]"}---which are $\Diff(M)$-invariant and represent \emph{physical} events/magnitudes. 
This has two natural dual implications.

The first is immediate: points and regions of $M$ are completely unphysical. 
\emph{Physical events and regions} are encoded as the $\Diff(M)$-invariant network of pointwise relations among the fields~$\upphi$. 
As \cite{Norton1989} says, ``the full force of the hole argument is directed against manifold [$M$] substantivalism": 
The~manifold $M$ is there only to bootstrap one's ability to erect a description of the relevant physics of fields, like a scaffold for a building. And like a scaffold, it is removed from the physical picture by the requirement of $\Diff(M)$-covariance of the field equations. 
The \emph{physical manifold of events} has no autonomous existence apart from the physical fields on which it supervenes. 

The second, which 
Earman and Norton called ``Leibniz equivalence", 
is that a whole $\Diff(M)$-orbit $\O_\upphi \subset \S$ represents one and the same physical solution, 
so that
\emph{physical field} degrees of freedom (d.o.f.) are not best represented by any single $\upphi \in \O_\upphi \subset \S$, but by a point (a $\Diff(M)$-equivalence class) $[\phi]\in \M_\S$ in the moduli space of solutions.
This is indeed the standard  view in physics. 
But the argument seems to further underlie that what is $\Diff(M)$-invariant and physical are the \emph{relations} among the d.o.f. of the fields $\upphi$, so that the physical fields d.o.f. mutually  co-define each other: 
The  d.o.f of the physical metric, i.e. of the gravitational field (redundantly described by $g$), are only well-defined in relation to the d.o.f. of matter and/or interaction fields (redundantly 
described by~$\upvarphi$).\footnote{
Considering a region of the Universe free of matter and interaction fields (without $\upvarphi$), one may submit that the d.o.f. of the physical metric/gravitational field are to be instantiated as invariant relations among those of the mathematical field $g$.}

The articulation of the hole and point-coincidence arguments makes a compelling case for a relationalist view of the \emph{physical} manifold of events, on  which 
the physical field d.o.f. ``live". 
In particular, 
the \emph{physical metric /gravitational field} is as autonomous an entity as it is possible in the general-relativistic framework,  physical field d.o.f. and the co-defining relations in which they participate being coextensive. 
Seeing that the gravitational field is the vessel of most of the essential properties usually attributed to spacetime, one may indeed consider that it embodies a substantive notion of spacetime, insofar as the ``substantival" notion gets suitably ``relationalized". 
This is not far from a family of views  called ``sophisticated substantivalism", see  \cite{Pooley2013-POOSAR,Stachel2014}.
\medskip

Let us add two comments before concluding this section.
First, {let us remind that Norton argued in \cite{Norton1989}}   
  that defining spacetime as ``$M$ + further structure" (``\emph{mpfs}") is a form of sophisticated substantivalism that may still be vulnerable to a hole-type argument if the ``further structure" has  symmetries, i.e.  a non-trivial Killing group. 
As we noted, the consensus in physics is to define spacetime as $(M, g)$. 
Furthermore, all exact solutions $g$ of Einstein's equations have a Killing group 
{K$(g)\defeq \left\{\ \psi \in \Diff(M)\, |\, \psi^*g=g \right\}$},
so substantivalism towards these symmetric spacetimes is undermined according to \cite{Norton1989}.
Yet, realistic physical solutions certainly have only approximate symmetries, so substantivalism towards these spacetimes would be possible. 
{Relatedly,}
the discussion of the physical significance of Killing symmetries by \cite{JTF-Ravera2024c} dissents from {Norton's} view by denying that the (sophisticated) substantivalist has to see Killing related models as physically distinct \emph{in the fully dynamically coupled regime}.\footnote{Killing symmetries of $g$ being symmetries of (the energy-momentum tensor of) $\upvarphi$ by Einstein equation $E(g, \upvarphi)=G(g)-\kappa T(g, \upvarphi)=0$.}

 {Next,}
 let us remark that in recent years there have been attempts to defuse the hole argument via what has been called ``formalist responses" \cite{sep-spacetime-holearg}. 
 It was ushered by \cite{Weatherall2018} defending  the claim that modern mathematical practice (category theory)  
 would short circuit the hole argument, while a complementary strategy was the attempt by  \cite{Halvorson-Manchak2022} to
 show that {a mathematical theorem prevents the hole argument to even be stated. 
 There have been a number of rebuttals, some of  which we essentially align with, notably \cite{Pooley-Read2021} and \cite{Menon-Read2023}. }
 
 {We shall give our own account of the issue and a detailed answer in a separate contribution \cite{Berghofer-et-al2027-A}. Here we restrict ourselves to a brief outline of our view of \cite{Weatherall2018} and \cite{Halvorson-Manchak2022}.
  Both works aim to leverage the fact that the category \textbf{Lor} of Lorentzian manifolds $(M, g)$ is the correct framework for GR,\footnote{{This may be granted for present purposes. However, note that here, as in most physics literature, one would rather argue that the appropriate category is (at least) the category \textbf{GR}$[\upphi]$, whose objects are $(M, \upphi)$, where $\upphi$ denotes a collection of fields (e.g. $\upphi=\{g, \varphi\}$ as above).}} so that the `correct' notion of transformation is that of \emph{isometry}: $\psi : (M, g) \rarrow (N ,h)$ is an isometry if $\psi \in \Diff(N, M)$ s.t. $h=\psi^*g$. The special case $N=M$ and $\psi \in \Diff(M)$ is immediate.
 The authors of \cite{Halvorson-Manchak2022} aim to block the hole argument by 
 arguing that a mathematical result by Geroch \cite{Geroch1969} proves that there can be no `hole isometry' in \textbf{Lor}, so that ``\emph{there is no mathematical fact that could underwrite the hole argument}". 
 We think this is incorrect, as Geroch's result just proves that there can be no \emph{hole automorphism}.
 The automorphism group is $\Aut(M, g)\defeq\{ \psi \in \Diff(M)\, |\, \psi^*g=g\}$, also denoted Iso$(M, g)$. Automorphisms are special isometries (from an object in \textbf{Lor} to \emph{itself}). Indeed, \cite{Geroch1969} does imply that $\Aut(M, g) \cap \Diff_c(M)=\{id_M\}$,  i.e. that there can be no compactly supported automorphism. However, the hole argument concerns $\Diff_c(M)$ rather than $\Aut(M, g)$.
 Turning to \cite{Weatherall2018}, the key aspect of the argument, in our understanding, is the claim that there is no room for the hole argument, since, from the perspective of the formalism, all objects in \textbf{Lor} related by isometries---including hole isometries given by $\Diff_c(M)$---are mathematically identical. 
 On this view, there is no standpoint from which $(M, g)$ and $(M, \psi^*g)$ can be regarded as physically distinct, since they have the same representational role. 
 We do not object to anything technical here (but may do so in \cite{Berghofer-et-al2027-A}), but simply observe that the latter statement, though reasonable, does not follow from the mathematics alone. Rather, it is a statement about the interpretation of the mathematics. 
 And when it comes to it, \cite{Weatherall2018} defends it in the spirit of the point-coincidence argument, asking which empirical differences an `observer' at a point would be able to notice (see section 3, especially footnote 20, where the relational undertones of the argument transpire). 
 So, it is tempting to understand \cite{Weatherall2018}'s argument as in some sense isomorphic to the articulation of the hole and point-coincidence arguments (previously summarized) but `done backwards'.\footnote{{We indeed find statements that are interpretable to that effect, notably: 
``\emph{This argument is just a recapitulation of the hole argument in different
terms [...] this new, different
sense in which $(M, g)$ and $(M, \psi^*g)$ turn out to be equivalent is precisely the sense we began with, namely, that they are isometric.}"}} 
In any case, we do not think that Weatherall's conclusions are inconsistent with ours. However, it is instructive to emphasize the difference in our respective approaches. Weatherall's formalist response maintains that the hole argument does not get off the ground because sound mathematical practice precludes its proper formulation. By contrast, 
we do not seek to block the hole argument by appealing to mathematical practice but embrace it and, as discussed below, by employing the Dressing Field Method we reformulate gauge theories in terms of dressed fields and spaces that are immune to it.
}

\medskip
{Indeed, if one sides, as we do, with the mainstream view that the hole argument stands in general,} 
its resolution by the point-coincidence  argument yields the physical insights summarized {earlier}.
These insights seem to be  those  Einstein took to be the final word of the general-relativistic framework.
In the note to the 15th edition of his book ``Relativity: The Special and the General Theory" \cite{Einstein1952-EINRTS-3}, he states that he added a new appendix, entitled ``Relativity and the Problem of Space", to present his ``\emph{views on the problem of space in general and on the gradual modifications of our ideas on space resulting from the influence of the relativistic view-point}", which he summarizes in the following few words: ``\emph{I~wished to show that space-time is not necessarily something to which one can ascribe a separate existence, independently of the actual objects of physical reality. Physical objects are not in space, but these objects are spatially extended. In this way the concept `empty space' loses its meaning.}"  
The last paragraphs of this appendix are worth quoting extensively:
\begin{quotation}
``We are now in a position to see how far the transition to the general theory of relativity modifies the concept of space. 
In accordance with classical mechanics and according to the special theory of relativity, space (space-time) has an existence independent of matter or field. 
In order to be able to describe at all that which fills up space and is dependent on the co-ordinates, 
space-time or the inertial system with its metrical properties must be thought of at once as existing, for otherwise
the description of `that which fills up space' would have no meaning.\footnote{``If we consider that which fills space (e.g. the field) to be removed, there still remains the metric space in accordance with (1) [the Minkowski metric], which would also determine the inertial behaviour of a test body introduced into it."} 
On the basis of the general theory of relativity, on the other hand, space as opposed to `what fills space', which is dependent on the co-ordinates, has no separate existence. 
Thus a pure gravitational field might have been described in terms of the $g_{ik}$ (as functions of the co-ordinates), by solution of the gravitational equations. 
If we imagine the gravitational field, i.e. the functions $g_{ik}$, to be removed, there does not remain a space of the type (1)
[i.e. of SR], but absolutely \emph{nothing}, and also no `topological space'. 
For the functions $g_{ik}$ describe not only the field, but at the same time also the topological and metrical structural properties of the manifold. 
A space of the type (1), judged from the stand-point of the general theory of relativity, is not a space without field, but a special case of the $g_{ik}$ field, for which---for the co-ordinate system used, which in itself has no objective significance---the functions $g_{ik}$ have values that do not depend on the co-ordinates. 
There is no such thing as an empty space, i.e. a space without field. Space-time does not claim existence on its own, but only as a structural quality of the field.

Thus Descartes was not so far from the truth when he believed he must exclude the existence of an empty space. The notion indeed appears absurd, as long as physical reality is seen exclusively in ponderable bodies. 
It requires the idea of the field as the representative of reality, in combination with the general principle of relativity, to show the true kernel of Descartes' idea; there exists no space `empty of field'." 
\end{quotation}

This is a masterful synthesis of the general-relativistic update of the  substantivalist \emph{vs} relationalist debate. 
Yet,~it~must be  reconsidered once more, this time in light of a framework that is a direct heir of GR (see \cite{ORaif1997}, \cite{Francois2021_II}, \cite{Francois2023-gRGFT-ontology}): Gauge Field Theory (GFT).
\enlargethispage{1\baselineskip}
This framework, of which Maxwell electromagnetism, the electroweak model and chromodynamics are models, 
offers
\emph{prima facie} a new view of spacetime. 
Taken together, general-relativistic field theory and GFT form the broad framework of \emph{general-Relativistic Gauge Field Theory} (gRGFT). 
Its mathematical foundation is the differential geometry of connections on fiber bundles, of which
we give a dense review in the next section.
It provides the necessary background to appreciate the notion of ``enriched spacetime" that naturally arises from gRGFT, and how  \emph{generalized} hole and point-coincidence arguments inform a realist view
towards such an enriched spacetime.

\section{Fiber bundles and basics of gauge theory}
\label{Fiber bundles and basics of gauge theory}

The geometry of
fiber bundles, and their connections, provides the kinematics of
gRGFTs. 
Yet, despite bundles being 
so fundamental to gauge theories,
only a few mathematical physicists endorsed some form of fiber
bundle \emph{realism}---see next section.
Here, we shall give a fairly technical review of the basics of the bundle geometry underpinning gRGFT, but with a more synthetic and careful logical progression than is typical.
This will provide the clear conceptual picture upon which we will heavily rely in our subsequent sections on the generalized hole and point-coincidence arguments, and on the Dressing Field Method (DFM). 
Readers who feel confident in their mastery of the subject can skip this section, referring back to it only to check mathematical notations.  

However, it is worth noting that while several authors such as \cite{Weatherall2016},  \cite{Jacobs2023} or \cite{Gomes2024} have recently emphasized \emph{associated vector bundles} when defending a form of bundle realism,\footnote{Mainly on account that sections of vector or affine bundles seem (to these authors at least) more immediate to interpret as physical fields. 
 \cite{Jacobs2023} e.g. thus proposing what the author calls ``\emph{associated bundle Platonism}".} we consider \emph{principal bundles} as the key spaces naturally ``hosting" all the physically relevant structures, their automorphism groups supplying both the diffeomorphisms and gauge groups of gRGFTs.\footnote{Anticipating the next section, they are thus the spaces on which the generalized hole and point-coincidence arguments operate.}
 Notably, this allows us to treat Yang-Mills {(YM)} theory and gravitational gauge theories in a unified manner, underpinned respectively by Ehresmann and Cartan connections.\footnote{It is well-known that the principal and associated bundles formulations are mutually translatable. For example, sections of associated bundles are  equivariant functions (or tensorial 0-forms) on a principal bundle, both mathematical notions adequately representing  matter fields. 
 The conceptual significance of this fact for bundle realism is commented upon in \cite{JTF-Ravera2024c}.
 We may yet observe that a principal bundle stands in relation with its associated bundles as a group stands in relation with its various representations: 
Arguably the former have some logical priority over, and ``control", the latter.
} 

Our presentation in this section is technically demanding but thorough and we hope it may serve the wider pedagogical purpose of introducing a broader audience to the mathematical foundations of classical gRGFT, enhancing its appreciation of the geometric origin of concepts most have likely encountered mainly in a field-theoretic context, which may  hopefully benefit philosophical discussions.
The~reader whose interest is piqued and who wishes to go further will profitably consult 
the book by \cite{Hamilton2017}, as well as the shorter text of \cite{Francois2021_II} (sections 1-4).

\medskip
A principal bundle is a smooth manifold $P$ supporting the smooth  right action of a Lie group $H$ called its \emph{structure group}, $P \times H \rarrow P$, $(p, h) \mapsto ph$.
The right action is also denoted $R_h\!: P \rarrow P$, $p\mapsto R_h p \defeq ph$.
The~orbits of the  action by $H$ are the \emph{fibers} of~$P$: by definition this action is free and transitive, so fibers are identical and isomorphic to $H$ \emph{as manifolds} but unlike $H$ have no group structure.   
 The~\emph{moduli} space of all fibers is itself a manifold $M$, called the \emph{base} of $P$: we have $P/H \defeq M$, and there is a canonical projection $\pi: P \rarrow M$, $\pi(ph)=\pi(p)=x$, i.e. s.t. $\pi \circ R_h =\pi$.
One often says that the point $x\in M$ has a fiber $P_{|x}$ \emph{attached} to it, or one calls $P_{|x}$ the fiber \emph{over} $x$. 
Sometimes a bundle $P$ 
with structure group $H$ is denoted $P\xrightarrow{\pi} M$. 
{To give two examples, in electroweak theory the structure group is $H=U(1)\times S\!U(2)$, and the fibers are compact spaces isomorphic to $S^1 \times S^3$. In the tetrad formulation of GR with spinors, the structure group is $H=S\!L(2,\CC)$---the double cover of orthochronous Lorentz group $S\!O^+(1,3)$---and the fibers are non-compact and isomorphic to $S^3 \times \RR^3$.}

The local structure of a bundle $P$ is trivial in that, given 
a finite region $U\subset M$,  $P_{|U}\simeq U \times H$. 
A bundle is said trivial if $P=M \times H$. 
Bundles are meant to generalize the trivial case. 
Local sections of $P$ over $U$ are  smooth maps $\s: U\rarrow P_{|U}$, $x\mapsto \sigma(x)$, s.t. $\pi \circ \s = \id_U$.
They provide  \emph{bundle charts/coordinates}. 

The~tangent bundle $TP$ of $P$ has a canonical \emph{vertical subbundle} $VP \subset TP$ whose sections, the ``vertical" vector fields, are tangent to the fibers of $P$; they may be defined as belonging to $\ker \pi_*$, where $\pi_*: TP \rarrow TM$ sends vectors on $P$ to vectors on $M$. 
The infinitesimal action of $H$ induces \emph{fundamental vertical vector fields}: For $X\in \LieH$ an element of the Lie algebra of $H$, let $c(\tau)\defeq ph(\tau) = p\exp{\{\tau X\}}$ be a curve in the fiber through $p=c(0)$, then $X_{|p}^v \defeq \tfrac{d}{d\tau} c(\tau)\big|_{\tau=0}$ is tangent to the fiber at $p$. 
 
\begin{figure}[t]
\begin{center}
\includegraphics[width=0.63\textwidth]{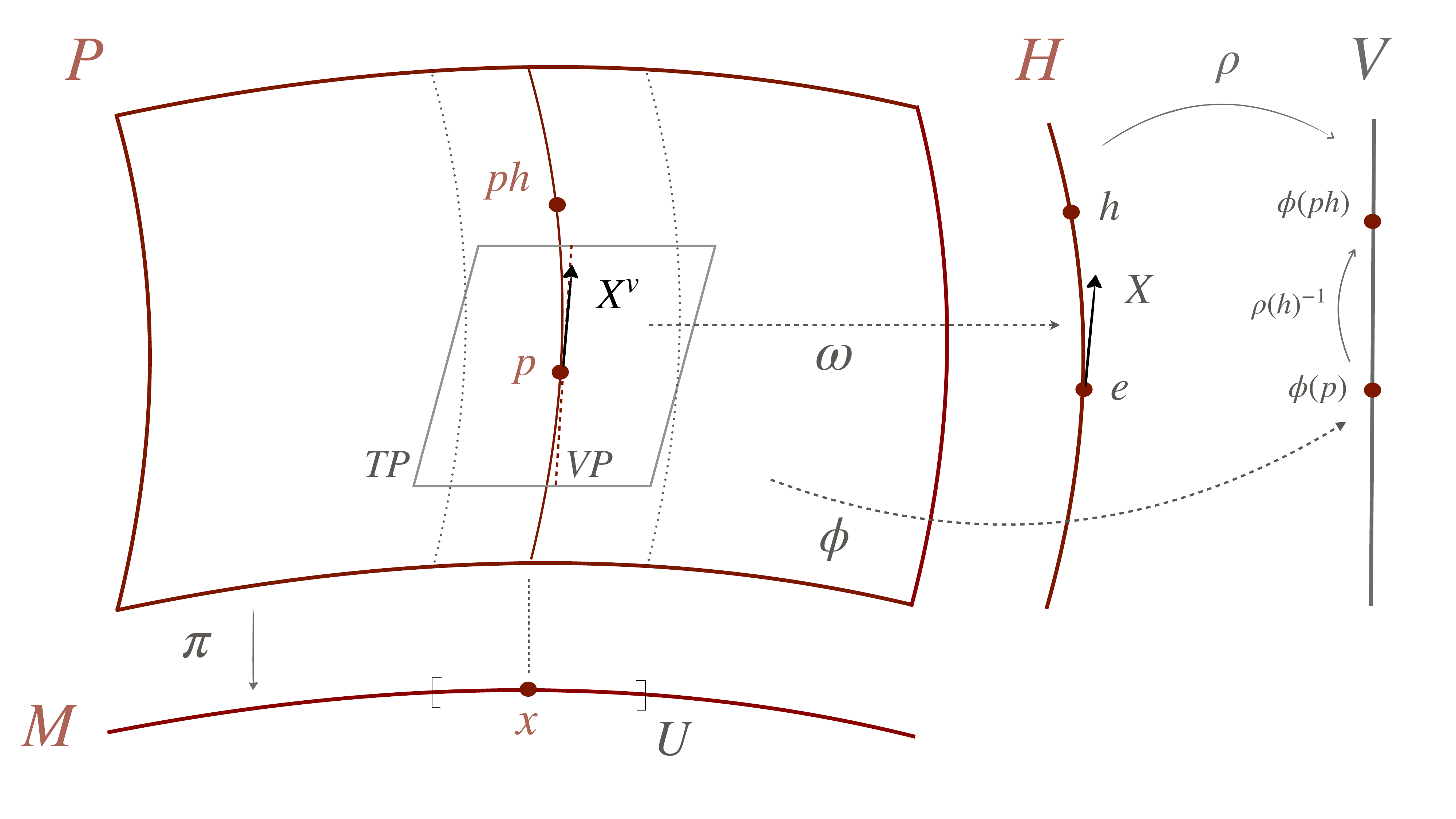}
\caption{{Schematic visualization of the principal bundle $P$, with structure group $H$, a vertical vector  $X^v$, a (principal) connection $\omega$, and tensorial $0$-form (matter field) $\phi$.}}
\label{bundlefig}
\end{center}
\end{figure}

Cartan calculus applies on $P$, so one may define the de Rham complex of differential forms $\big( \Omega^\bullet(P), d \big)$ with $d$ the de Rham, or exterior, derivative. 
Forms that vanish when evaluated on vertical vector fields are said \emph{horizontal}. 
The pullback of a form $\alpha$ by the right action of $H$ defines its ``equivariance": $R^*_h\!: T_{ph}P \rarrow T_pP$, $\alpha_{|ph}\mapsto R_h^*\alpha_{|ph}$ for  $h\in H$. 
It~describes how $\alpha$ changes when moved along the fibers via $H$. 
Given a representation of $H$, $\rho:H \rarrow GL(V)$, $h\mapsto \rho(h)$, with $V$ a vector space, one  defines
the space of representation-valued \emph{equivariant forms} as those  whose equivariance is simple and controlled by the representation,  $\Omega^\bullet_{\text{eq}}(P, V):=\big\{\, \alpha \in \Omega^\bullet(P, V)\ |\  R_h^*\alpha_{|ph}= \rho(h\-) \alpha_{|p}) \,\big\}$. 
Forms that are both horizontal and equivariant are said \emph{tensorial}, for reasons soon disclosed.
Forms with trivial equivariance, i.e. s.t. $R^*\alpha_{|ph}=\alpha_{|p}$, are said \emph{invariant}. 
Forms that are both horizontal and invariant are said \emph{basic}: they are important as they induce forms on $M$, hence their name, and may be defined as belonging to  $\im\pi^*$, the image of  $\pi^* : \Omega^\bullet(M) \rarrow \Omega^\bullet(P)$, $\beta \mapsto \alpha\defeq \pi^*\beta$. 
{Some of the key structures described above can be displayed schematically as in Fig. \ref{bundlefig}.}
\medskip

The  maximal group of transformations of  the bundle $P$ is its group of automorphisms,
\begin{align}
    \Aut(P):=\big\{ \Xi \in \Diff(P)\ |\ \Xi(ph)=\Xi(p)h \big\},
\end{align}
i.e. its elements are those diffeomorphisms of $P$ that respect its fibration structure by sending fibers to fibers, and thus induces well-defined diffeomorphisms of the base $M$: we have  the natural surjection $\Aut(P) \rarrow \Diff(M)$. 
The subgroup of vertical automorphisms is $\Aut_v(P):=\big\{ \Xi \in \Aut(P)\ |\ \pi \circ \Xi=\pi\big\}$, i.e. its elements are those automorphisms of $P$ that induce the identity transformation $\id_M$ on $M$. 
It~is a \emph{normal} subgroup of $\Aut(P)$, which is noted $\Aut_v(P) \triangleleft \Aut(P)$, and implies that their quotient $\Aut(P)/\!\Aut_v(P)$ is itself a group, which is isomorphic to $\Diff(M)$.\footnote{{Notice that our view is ``top-down": we regard the bundle $P$, together with its automorphism group $\Aut(P)$, as the \emph{primary structure}, while the base manifold $M$ is a secondary, derived structure, whose automorphism group $\Diff(M)$ is induced by $\Aut(P)$. 
 This is close in spirit to the `gauge natural bundles' mathematical school, as reflected e.g. in 
\cite{Kolar-Michor-Slovak} or \cite{Cap-Slovak09}. 
And it fits the spirit of the `bundle realist' stance we defend here in the sense that, on a realist/substantivalist view, one expects a single bundle space with its given topology to be physically realized, and to be represented mathematically by a given $P$ and its $\Aut(P)$-class.
It~is a distinct, ``bottom-up", attitude to consider $M$ primary, with its automorphism group $\Diff(M)^*$, and then to ask the classification question: ``How many topologically distinct $H$-bundles could exist above $M$?", and the follow-up ``Which subgroup of $\Diff(M)^*$ lifts to $\Aut(P)$ for a $P$ in a given topology class?". The answer to the latter being $\Diff(M)=\left\{\psi \in\Diff(M)^*\,|\, \psi^* P \simeq P \right\}$---where $\psi^* P$ means the pullback bundle. We thank a referee for suggesting clarifications on this point. 
The classification question may become especially relevant to a bundle realist view when considering (in direct analogy with quantum gravity) the issue of quantum fluctuations or tunneling between  physical bundles with distinct topologies, or the issue of how these contribute to a path integral. We set these questions aside here, but will return to them in future work.}
}
All this is synthesized in the \emph{short exact sequence} of groups associated to a bundle $P$\,:
\begin{align}
\label{SES-P}
\id_P\ \rarrow\  \Aut_v(P) \simeq\! \H \ \xrightarrow{\triangleleft}\  \Aut(P)  \ \xrightarrow{} \  \Diff(M)\  \rarrow \ \id_M.
\end{align}
It also features the fact that
the subgroup $\Aut_v(P)$ is isomorphic to the \emph{gauge group} of $P$, which is defined as $\H:= \big\{ \upgamma:P \rarrow H\ |\ R^*_h \upgamma =h\- \upgamma h \big\}$, i.e. its elements are s.t. $\upgamma(ph)=h\-\upgamma(p)h$,  the isomorphism being $\Xi(p)=p\upgamma(p)$.\footnote{Beware not to confuse the notion $\H$ for the gauge group of $P$ with the familiar notation for Hilbert spaces, $\mathscr{H}$.} 
{In~electroweak theory we have $\H=\U(1)\times \SU(2)$, in GR with spinors we have $\H=\SL(2, \CC)$.}

The action by pullback of $\Aut_v(P)$ on a form $\alpha$ defines its  \mbox{\emph{gauge transformation}}, 
$\Xi^*\!: \Omega^\bullet(P)_{|\Xi(p)} \rarrow \Omega^\bullet(P)_{|p}$, $\alpha \mapsto \Xi^*\alpha \rdefeq \alpha^\upgamma$. 
The defined notation $\alpha^\upgamma$ is justified by the above isomorphism, by which  the gauge transform  is expressible in terms of  $\upgamma \in\H$ generating  $\Xi \in \Aut_v(P)$.
It describes how $\alpha$ changes when moved along the fibers via $\Aut_v(P)\simeq \H$. 
For tensorial forms,  gauge transformations are entirely controlled by their equivariance, thus by the representation, 
$\alpha^\upgamma = \rho(\upgamma)\-\alpha$: 
they are  then homogeneous,``gauge-tensorial", hence their name. 
Remark that, since $\upeta \in \H$ is also  $\upeta \in \Omega^0_\text{tens}(P, H)$, the gauge transformation of a gauge group element is: $\upeta^\upgamma =\upgamma\- \upeta \upgamma$ for $\upeta, \upgamma \in \H$. 
For~basic forms, gauge transformations are trivial $\alpha^\upgamma =\alpha$, as  expected for a form inducing a well-defined object on $M$ where the action of $\Aut_v(P)$ is trivial. 
Both tensorial and basic forms are essential for gauge field theory: the latter capture gauge-invariant (``physical") d.o.f., while e.g. tensorial 0-form $\phi$ 
describe \emph{matter fields}.
\medskip

Up to now, all structures described on $P$ are canonical---even the representation $(\rho, V)$ ingredients, which arguably come along with the structure group.
The first and most important non-canonical object to be described is a \emph{connection 1-form},  $\omega \in \Omega^1_\text{eq}(\Omega, \LieH)$:  by definition its equivariance is controlled by the adjoint representation $\Ad: H \rarrow GL(\LieH)$, so that $R^*_h \omega = \Ad(h\-)\,\omega \defeq  h\- \omega\, h$, and it is required to satisfy $\omega(X^v)=X \in \LieH$ on fundamental vertical vector fields. 
These two properties imply: 
First, that the space of connections is an affine space modeled on the vector space $\Omega^1_\text{tens}(P, \LieH)$, meaning that while addition of connections is impossible, still $\omega'- \omega= \alpha \in \Omega_\text{tens}^1(P, \LieH)$.
Secondly, that the gauge transformation of a connection is $\omega^\upgamma= \upgamma\- \omega \upgamma + \upgamma\-d\upgamma$, i.e. it is ``gauge pseudo-tensorial".
The above is known as an Ehresmann (or principal) connection. It is the one underlying ``Yang-Mills type" GFT.

Gauge theories of gravity are based on \emph{Cartan connections},\footnote{{By  ``gauge" theories of gravity we mean those with  an `internal' gauge group in addition to $\Diff(M)$. 
These include
all kinds of ``tetrad formulations" of GR, i.e. with   $\H=\SO(1,3)$ (and $\LieG=\so(1,3) \oplus \RR^4$), which are necessary to describe the coupling of gravity to spinorial matter fields (fermions)---for which $\rho: H=S\!O(1,3)\rarrow S\!L(2,\CC)$---and featuring  non-vanishing torsion. 
It also encompasses more speculative gauge theories of gravity for which $\H \supset \SO(1,3)$---or $\H\supset \SL(2,\CC)$. 
A notable example is the historical ``first" gauge theory, Weyl's unified theory of 1918-1919, which relies on the restricted conformal group $C\!O(1,3) = S\!O(1,3) \times W$, with  $W=\RR/\{0\}$  the Weyl group of rescaling; it is based on Cartan-Weyl geometry.
Another is conformal gravity, based on the conformal Lie algebra $\LieG=\so(2,4)$, for which $H= C\!O(1,3) \ltimes \RR^{4*}$, with $\RR^{4*}$ the group of special conformal transformations; it is grounded in  conformal Cartan  geometry, which is also the root of twistorial gravity and twistor theory---see \cite{Attard-Francois2016_II}. 
Conformal Cartan geometry, together with projective Cartan geometry (the latter underlying projective gravity theories, and directly relevant to 2D conformal field theory), are prototypical examples of parabolic Cartan geometries---see \cite{Cap-Slovak09}.
Supergravity theories are similarly  based (or ought to be) on Cartan supergeometry; see the review by \cite{JTF-Ravera2024review}.}}
{which} 
are 1-forms $\varpi \in \Omega^1_\text{eq}(P, \LieG)$, where $\LieG \supset \LieH$ and $\LieG/\LieH=V$ 
supports a left-action of $H$,
satisfying the same defining properties as $\omega$, plus a third distinctive one: for all $p\in P$, $\varpi: T_pP \rarrow \LieG$ is a linear isomorphism. 
This is the key feature of Cartan geometry, making it the foundation of gauge gravity, as it means that the geometry of $P$ encodes that of $M$.
In many cases, it implies that a Cartan connection splits as $\varpi =\omega \oplus \theta$, where $\omega$ is an Ehresmann connection and $\theta \in \Omega^1_\text{tens}(P, V)$ is a \emph{soldering form}.
The~properties of Cartan connections imply, first that they form an affine space modeled on the vector space $\Omega^1_\text{tens}(P, \LieG)$, and secondly that their gauge transformation is $\varpi^\upgamma= \upgamma\- \varpi \upgamma + \upgamma\-d\upgamma$, which splits into $\omega^\upgamma= \upgamma\- \omega \upgamma + \upgamma\-d\upgamma\,$ and $\,\theta^\gamma =\upgamma\- \theta$. 

The reason to introduce an Ehresmann connection is to obtain a first-order differential operator on $\Omega^\bullet_\text{tens}(P, V)$: this space is not preserved by $d$, so one defines the \emph{covariant derivative} $D\_ \defeq d\_ + \rho_*(\omega)\_: \Omega_\text{tens}^\bullet(P, V) \rarrow \Omega_\text{tens}^{\bullet+1}(P, V)$, $\alpha \mapsto D\alpha$, where $\rho_*:\LieH \rarrow \mathfrak{gl}(V)$. 
Which means in particular that $\alpha$ and $D\alpha$ have the same gauge transformations, their ``gauge-tensoriality" is preserved: $\alpha^\upgamma=\rho(\upgamma)\-\alpha$ and $(D\alpha)^\upgamma=\rho(\upgamma)\-D\alpha$.
As applied to $\alpha = \phi \in \Omega^0_\text{tens}(P, V)$, this is the geometric underpinning of the \emph{Gauge Principle}, or \emph{gauge argument}.
{It further implies, let us stress,  that a connection introduces a standard against which ``internal constancy" is determined; this is what it means for some $\alpha$ to be \emph{covariantly constant}, $D\alpha=0$. 
This property, which, notice, is a
\emph{gauge-invariant} one, determines the ``parallel transport" of $\alpha$ across the internal geometry of $P$, i.e. from fiber to fiber: it determines ``free" motion in $P$}.\footnote{{This, with Cartan geometry, encompasses in particular parallel transport on pseudo-Riemannian manifolds, i.e. geodesic motion in GR.}}

The \emph{curvature 2-form} of a connection is  $\Omega=d\omega+ \tfrac{1}{2}[\omega, \omega] \in \Omega^2_\text{tens}(P, \LieH)$. Its gauge transformation is thus $\Omega^\upgamma = \upgamma\-\Omega \upgamma$, it is a ``gauge tensor". Its covariant derivative is trivial, which just expresses the Bianchi identity: $D\Omega=d\Omega+[\omega, \Omega]=0$, where $[\omega, \_]=\ad(\omega)\_=\Ad_*(\omega)\_$\,. 
One also easily proves that $D \circ D = \rho_*(\Omega)$ on $\Omega^\bullet_\text{tens}(P, V)$.  
The curvature of a Cartan connection is  $\b\Omega=d\varpi+ \tfrac{1}{2}[\varpi, \varpi] \in \Omega^2_\text{tens}(P, \LieG)$, splitting as $\b\Omega=\Omega \oplus \Theta$, where $\Theta=D\theta$ is the torsion 2-form of $\varpi$. Its gauge transformation is 
$\b\Omega^\upgamma= \upgamma\- \b\Omega\upgamma $, which splits into $\Omega^\upgamma= \upgamma\- \Omega \upgamma$ and $\Theta^\gamma =\upgamma\- \Theta$.
It satisfies the Bianchi identity: $\b D\b\Omega\defeq d\b\Omega+[\varpi, \b\Omega]=0$. 
One may check e.g. that the Cartan geometry based on $(\LieG, \LieH)$ the Poincaré and Lorentz Lie algebras, is just Lorentzian geometry, $\H$ being then the Lorentz gauge group. 

\medskip

General-relativistic Gauge Field Theory (gRGFT) is written on  the base manifold $M$ of a bundle $P$ rather than on $P$ itself, or even on a finite region $U\subset M$ over which the bundle is trivializable, $P_{|U}\simeq U \times H$. 
The fields of gRGFT  are the \emph{local representatives} on $U$ of global objects ``living" on $P$, the former being the pullback of the latter via any local section: given $\s:U \rarrow P_{|U}$, we have $\s^*\!\!:\Omega^\bullet(P_{|U}) \rarrow \Omega^\bullet(U)$,  $\beta \mapsto \s^* \beta \rdefeq b$. 
Said otherwise, the local representatives $b$ are the ``bundle coordinate" versions of the intrinsic global  objects $\beta$.
 
Then, the local representatives
of an Ehresmann connection $\omega$ and its curvature $\Omega$ are 
$A\defeq \s^* \omega \in\Omega^1 (U, \LieH)$ 
and $F\defeq \s^*\Omega = dA + \tfrac{1}{2}[A, A] \in\Omega^2 (U, \LieH)$, which represent respectively a \emph{gauge potential} and its \emph{field strength}.
{In electromagnetism, for $H=U(1)$, this gives $F=dA$, which in coordinates is $F=\tfrac{1}{2} F_{\mu\nu}\, dx^{\,\mu} \w dx^\nu$, with $F_{\mu\nu}$ the Maxwell-Faraday tensor.
The non-Abelian generalization yields in components the Yang-Mills field strength $F^a_{\mu\nu}$,  the Lie algebra index $a$ being the internal ``color" index.}
The local representatives of a tensorial form and its covariant derivative, $\alpha, D\alpha \in \Omega^\bullet_\text{tens} (P, V)$, are $a\defeq \s^*\alpha \in \Omega^\bullet (U, V)$ and $Da\defeq s^*(D\alpha)=da+ \rho_*(A)a$. 
From which we have that $F$ satisfies the (local) Bianchi identity $DF=dF +[A, F]=0$.
In particular, the local representative of $\phi \in \Omega^0_\text{tens}(P, V)$ is a \emph{matter field} $\vphi\defeq \s^* \phi$, 
and $D\vphi=d\vphi+\rho_*(A)\vphi$ is just its  \emph{minimal coupling} to the gauge potential---usually arrived at via the heuristic of the ``gauge argument"~in~GFT.

The local representative of a Cartan connection $\b A\defeq \s^* \varpi \in\Omega^1 (U, \LieG)$ represents the (generalised) gravitational gauge potential. 
It splits as $\b A = A + e$,  where $e\defeq \s^*\theta \in \Omega^1 (U, V)$ is the \emph{vielbein} 1-form, {which is in coordinates $e=e^a={e^a}_\mu\, dx^{\,\mu}$, with ${e^a}_\mu$ the \emph{co-tetrad} field, and where $a$ is an ``internal" $V$-index (e.g., a Minkowski index)}.
Given a (non-degenerate symmetric) bilinear form {$\langle\ , \rangle: V \times V \rarrow \RR$, $(v, w) \mapsto \langle v, w \rangle$, the vielbein $e$ induces a metric on $M$ by $g\defeq \langle\ , \rangle \circ e: TM \times TM \rarrow \RR$, $({\mathscr X}, {\mathscr Y}) \mapsto g({\mathscr X}, {\mathscr Y})\defeq \langle e({\mathscr X}), e({\mathscr Y})\rangle$,} in coordinates; $g_{\mu\nu}=\eta_{ab}{e^a}_\mu{e^b}_\nu$. 
The~gravitational field strength is $\b F\!\defeq \s^*\b \Omega = F + T$, with $T\!\defeq \s^*\Theta=De \in \Omega^2 (U, V)$ the torsion, and s.t. $\b D\b F=0$.
{For $\LieG$ the Poincaré algebra, $A$ is the Lorentz/spin connection and $e$ the tetrad field, both featuring in Einstein-Cartan gravity; then, $F=R$ is the Riemann 2-form, in components $R=\tfrac{1}{2} {R^a}_{b,\, \mu\nu} \,dx^{\,\mu}\w dx^\nu$, with ${R^a}_{b,\, \mu\nu}$ the Riemann tensor, and similarly $T$ in components is ${T^a}_{\mu\nu}$, the torsion tensor.}
\medskip

General-relativistic physics requires covariance under $\Diff(M)$, while Gauge Field Theory requires invariance under the gauge group $\H$, or rather the local gauge group on $M$ i.e. 
\begin{equation}
\label{loc-gaugr-grp}
\H_\text{loc}:=\big\{  \gamma=\s^*\upgamma, \upgamma \in \H \ |\ \eta^\gamma=\gamma\- \eta \gamma \big\}.
\end{equation}
{The defining property $\eta^\gamma=\gamma\- \eta \gamma$ reflects that of the elements of the gauge group $\H$---as discussed after eq. \eqref{SES-P}; it means that a local gauge group element $\gamma$ acts on another $\eta$ in the prescribed way.}
The~full group of local symmetries of gRGFT is thus $\Diff(M)\ltimes \H_\text{loc}$ which is just the local description of $\Aut(P)$. {The semi-direct product being 
\begin{align}
\label{semi-dir-prod}
(\psi', \gamma')\bs{\cdot}(\psi, \gamma)= \big(\psi' \circ \psi, \, \gamma' \, [\gamma \circ {\psi'}^{-1}] \big),
\end{align}
where $\circ$ is the composition of maps and the product in $\Diff(M)$.}
The~group $\Diff(M)$ acts on any fields or forms by pullback: 
Taking from now on as our elementary variables the Yang-Mills~and/or gravitational gauge potentials and matter fields, 
$\upphi= \lbrace{A, \vphi ,  e/g, \rbrace}$,  
for $\psi \in \Diff(M)$ we have  $\upphi^\psi\defeq \psi^* \upphi$, i.e. 
\begin{align}
\label{Diff-trsf-fields}
A^\psi \defeq \psi^* A, 
\quad 
\vphi^\psi  \defeq \psi^* \vphi, 
\quad \text{and} \quad
e^\psi\defeq \psi^*e \ \rarrow\ 
g^\psi\defeq \psi^*g.
\end{align}
We write similarly for minimal couplings $D\vphi$ and for the Yang-Mills and/or gravitational field strengths $F$, $\b F$.
The~action of the gauge group $\H_\text{loc}$ on $b$ is obtained by pullback of the result of the action of $\H$ on $\beta$:
$b^\gamma \defeq \s^*(\beta^\upgamma)$.
So, for $\gamma \in \H_\text{loc}$  we have that $\upphi^\gamma$ is:
\begin{align}
\label{GT-trsf-fields}
A^\gamma = \gamma\-  A \gamma  + \gamma\-d\gamma, 
\quad 
\vphi^\gamma  \defeq \rho(\gamma)\-\vphi, 
\quad \text{and} \quad
e^\gamma\defeq \gamma\-e \ \rarrow\ 
g^\gamma\defeq g.
\end{align}
Similarly, or consequently, we have $(D\vphi)^\gamma = \rho(\gamma)\-D\vphi$ and $F^\gamma=\gamma\- F \gamma$, as well as  $\b F^\gamma=\gamma\- \b F \gamma$ with $T^\gamma =\gamma\- T$. 
Consider that in \eqref{GT-trsf-fields}, one should  typically understand that $\H_\text{loc}=\H_\text{loc}^{\text{\tiny {YM}}} \times \H_\text{loc}^\text{grav}$: 
The Yang-Mills gauge group $\H_\text{loc}^{\text{\tiny {YM}}}$ acting only on  the Yang-Mills gauge potential $A=A'$ and  matter fields $\vphi$, 
but trivially on the gravitational potential $\b A=A+e $ (thus~on $g$), 
while the gravitational gauge group $\H_\text{loc}^\text{grav}$ acts trivially on a Yang-Mills potential $A=A'$, 
but non-trivially on matter fields $\vphi$, which are  \emph{spinorial} representations of $\H_\text{loc}^\text{grav}$, and on gravitational potential $\b A =A +e$ (yet still trivially on $g$ if $\langle\ , \rangle$ is $\H_\text{loc}$-invariant).\footnote{In established physics $\H_\text{loc}^\text{grav}$ is the Lorentz gauge group $\SO(1,3)_\text{loc}$, so $g$ is indeed invariant as $\langle\ , \rangle$ is the Minkowski form on $V=\RR^4$. 
In~speculative theories the action of $\H_\text{loc}^\text{grav}$ on $g$ may be  non-trivial:  e.g. in conformal geometry/gravity, it induces a conformal transformation, $g\mapsto g^\gamma = z^2g$ for $z\in \W= C^\infty(U, \RR/\{0\})$. 
}
{
The combined action by $(\psi, \gamma) \in \Diff \ltimes \H_\text{loc}$ is $\upphi^{(\psi, \gamma)}:=\psi^*(\upphi^\gamma)$.\footnote{{Which means that, in writing $(\psi, \gamma)$, while the argument of $\psi$ is (say) $x$, that of $\gamma$ is $\psi(x)=:x'$. Iterating, in $(\psi', \gamma')$, one evaluates $\psi'$ at $x'$ and  $\gamma'$ at $\psi'(x')=\psi'\circ \psi(x)$.
This is consistent with the product rule \eqref{semi-dir-prod}, where if $\psi'\circ \psi$ is evaluated at $x$ then $\gamma' \, [\gamma \circ {\psi'}^{-1}] $ is evaluated at $\psi'\circ\psi (x)$, yielding $\gamma'(\psi'(x'))\, \gamma(\psi(x))$.}}
}

A general-relativistic gauge field theory is defined by a Lagrangian functional on the   field space $\Phi=\{A, \vphi, e/g \}$, 
\begin{equation}
\begin{aligned}
 L: \ \Phi\   &\rarrow\  \Omega^m(U, \RR),\\
 \upphi 
 &\mapsto L(\upphi)=L(A, \vphi, e/g),
\end{aligned}
\end{equation}
with $m=$ dim$M$, i.e. the Lagrangian is a top form on $M$. 
As said above, the Lagrangian is required to be covariant under $\Diff(M)$ and quasi-invariant---i.e.  invariant up to a boundary term---under $\H_\text{loc}$:
\begin{align}
\label{Sym-L}
L(\upphi)^\psi\defeq L(\upphi^\psi)=\psi^* L(\upphi)
\qquad \text{and} \qquad 
L(\upphi)^\gamma\defeq L(\upphi^\gamma) = L(\upphi) + db (\gamma; \upphi).
\end{align}
{For example, the YM Lagrangian  $L_\text{YM}(A) = \kappa \Tr(F \w *F)$, with $\kappa$ a constant (and where $*$ is the Hodge operator), is strictly invariant under $\H_\text{loc}^{\text{\tiny {YM}}}=\SU(n)_\text{loc}$. The Einstein-Hilbert-Cartan Lagrangian $L_\text{YM}(\b A)= \kappa\ R^{ab} \w e^c\w e^d\, \epsilon_{abcd}$ (with $\epsilon_{abcd}$ the Levi-Civita tensor) is strictly invariant under $\H_\text{loc}^\text{grav}=\SO(1,3)_\text{loc}$. }
This ensures that the field equations $E(\upphi)=E(A, \vphi, g)=0$ are covariant under $\Diff(M)\ltimes \H_\text{loc}$, that is:
\begin{align}
\label{cov-E}
E(\upphi)^\psi=E(\upphi^\psi)=\psi^*E(\upphi)=0
\qquad \text{and} \qquad
E(\upphi)^\gamma=E(\upphi^\gamma)=\uprho(\gamma)\-E(\upphi)=0,
\end{align}
where $\uprho=\{\rho, \Ad, \ldots \}$ denotes the various representations of $H$ (thus $\H$), including the trivial one, to which the various fields under consideration belong to.
\medskip

In gRGFT,  the base manifold $M$ is  typically regarded as representing the manifold of (spatio-temporal) events,
and $(M, g)$ representing spacetime. 
On that view, the bundle structure $P$ ``over" $M$ may be understood as meaning that to each spatio-temporal events $x\in M$ there are ``attached" identical fibers $P_{|x}$ that represent an internal, i.e. non-spatio-temporal, space. 
The bundle $P$ thus represents an \emph{enriched} manifold of events whose ``points" are not structureless, as in general-relativistic physics, but endowed with an \emph{internal} structure.
The various gauge fields of gRGFT then have access to (``probe") this enriched space, and have thus both  spatiotemporal and internal d.o.f.

Yet, as reminded in section \ref{The hole argument}, the hole argument 
 challenges  ``naive" substantivalism regarding both the \mbox{manifold} of events $M$ and spacetime $(M, g)$, and  when paired with the point-coincidence argument, invites a more \mbox{subtle} view whereby the true manifold of physical spatio-temporal events arises in the network of relations among \mbox{fundamental} fields, with the physical metric/gravitational field playing a co-defining role. 
Physical spacetime  is then represented in the general-relativistic framework in a way that is strictly concordant with neither the usual relational nor \mbox{substantival} views.
It is then natural to expect, given that in gRGFT $M$ arises as the quotient of the richer space~$P$, that \emph{\mbox{generalized} hole} and \emph{point-coincidence arguments} similarly
contribute to a better understanding of possible realist positions about the principal bundle.
\medskip

\section{The generalized hole argument in general-relativistic gauge field theory}\label{The generalized hole argument}

By bundle realism, we understand the view that the physical manifold of events has the structure of a fiber bundle; 
{in addition to physical ``spatio-temporal directions" there are physical ``internal directions", the latter being of a distinct ``nature" from the former and non-reducible to them.}
{Or, said otherwise,
each spatio-temporal point has an internal structure, 
a physical internal space,  constituting a fiber of the total physical bundle space.} 
{The ``motion" within which is controlled by its connection structure: matter fields (represented by tensorial forms) interact with  gauge fields (represented by connections) via minimal coupling, which implements the \emph{parallel transport}---i.e. ``free" motion---of the matter fields within the (spatio-temporal and internal) geometry of the physical bundle.\footnote{{Notice that accepting the physicality of the internal space does not imply accepting that one must think of it as just another \emph{spatial} extension, i.e. as d.o.f. of the same `nature' as those of the spatio-temporal order, as one would do in Kaluza-Klein (KK) type theories. In the latter, the higher-dimensional space-time indeed has the structure of a bundle space, as there is by hypothesis a privileged fibration along  the ``extra" (compact) dimensions; but  it is endowed with a metric structure---decomposed into a standard metric, gauge fields  off-diagonal, plus extra scalars (`dilatons' or `radions', unphysical as far as is known)---which 1) enforces the idea that the extra dimensions have the same `spatial' quality as the extended ones, and 2) is a less general structure than a connection. 
In standard (gR)GFT, the connection or gauge field `on' the physical bundle space  1) is just general enough to implement parallel transport across the internal geometry, from fiber to fiber, and 2) respects the distinctive nature of the internal direction as non-reducible to `spatio-temporal' qualities. 
A KK type view of an enriched physical spacetime is thus rather conservative, and less radical ontologically than that underpinned by (gR)GFT; it is thus remarkable that the latter has considerable empirical support, while the former has none.}}}

Given that our most fundamental physical theories are models of the gRGFT framework (and quantization thereof, in the case of the SM), 
one 
might expect that questions about the ontological status of fiber bundles would be among the most important topics discussed in the philosophy of physics.
Yet, the literature on this is surprisingly sparse. 
Physicists and mathematical physicists may occasionally stress the technical importance of bundles, 
for example:
\begin{quotation}
``It is a widely held view among mathematicians that the fiber bundle is a natural geometrical concept. Since gauge fields, including in particular the electromagnetic field, are [connections on] fiber bundles, all gauge fields are thus based on geometry. To us it is remarkable that a geometrical concept formulated without reference to physics should turn out to be exactly the basis of one, and indeed maybe all, of the fundamental interactions of the physical world." \cite{WuYang}
\end{quotation}

\begin{quotation}
``For me, a gauge theory is any physical theory of a dynamic variable which, at the classical level, may be identified with a connection on a principal bundle." \cite{Trautman1980}
\end{quotation}

\begin{quotation}
``What is gauge theory? It is not an overstatement to say that gauge theory is ultimately the theory of \emph{principal bundles} 
[...]. The fundamental geometric object in a gauge theory is a principal bundle over spacetime [...]."
\cite{Hamilton2017}
\end{quotation}
But they almost invariably stop short of truly expressing any ontological commitment. 
In philosophy of physics, while the literature on the significance of gauge symmetries ($\H_\text{loc}$) has  grown considerably  in the last 25 years, making it a well-established topic, 
the ontological status of the spaces of which they are automorphisms, bundle spaces, have comparatively received little attention.\footnote{Which is not to say that bundles are not often acknowledged in this literature as important mathematical structures. This is a curious situation; comparable to a fictitious one in which the community would have extensively discussed the physical meaning of the group of diffeomorphism $\Diff(M)$, but neglected the issue of the ontological status of the manifold $M$.}
Exceptions we know of, which at least raise or strongly hint at the question, being \cite{Stachel1986,Lyre1999, Guttmann-Lyre2000,Lyre2004,Lyre2004book,Healey2001,Nounou2003,Healey2007,Maudlin2007,Arntzenius2012,Dewar2019,Catren2022,Jacobs2023,Gomes2024,JacobsRead}.
One may consult \cite{Stachel2014}'s extensive defense of the relevance of the bundle viewpoint both in GFT and GR.

We are now in a position to discuss substantivalist and relationalist approaches to the principal fiber bundle. 
Echoing the classical opposition, the former{, taking the mathematical formalism at face value, }would contend that $P$ represents a physical entity, the enriched manifold of events, 
which has an existence autonomous from the various gauge fields inhabiting it{, and thus acts as a `container' for them}. The latter would deny this claim. 
As far as we can tell, \cite{Maudlin2007} and \cite{Arntzenius2012} lean toward a bundle substantivalist view.
In the physicists' camp,  Ne'eman is a rare exception on the record insisting that ``physics selects the \emph{realist} or \emph{substantivist} view" \cite{Neeman1996}, with \cite{Lyre2004} reporting that in personal conversation Ne'eman explicitly characterized his position as a bundle substantivalism.
On the contrary, \cite{Lyre1999} and \cite{Healey2001} would appear to be at least anti-substantivalists; 
a position they rest on the fact that bundle substantivalism is vulnerable to an ``internal hole argument".  
\cite{Lyre1999} saying that ``\emph{there exists a straightforward extension of the spacetime manifold hole argument to a generalized bundle space hole argument}", while \cite{Healey2001} argues that ``\emph{fiber bundle substantivalism [...] is subject to an analogue of the ``hole" argument against space-time substantivalism}".

We  argue below for a form of sophisticated bundle substantivalism,
not unlike the ``sophistication" of \cite{Jacobs2023}, though our view is more explicitly and essentially relational---and the end result of an interpretive process, rather than its starting point. 
Let us first notice that, 
{at least in  traditions  operating within a form of scientific realism},\footnote{{
Which, roughly construed, is the stance that we should adopt a positive epistemic attitude toward scientific claims and take scientific theories and concepts at “face value.” More specifically, scientific realism is often analyzed along three dimensions of realist commitment. Ontologically, realists are committed to the existence of a mind-independent world investigated by the sciences. Semantically, they endorse a literal interpretation of scientific theories. Epistemologically, realists champion the view that science provides us with knowledge of the world. When we say that the empirical success of GR is typically taken to support a realist view of spacetime, we have in mind the ``No Miracles Argument", according to which the most plausible explanation of the high degree of experimental accuracy of a theory is that its central concepts and structural features track, at least approximately, the way the relevant domain of reality is structured. Of course, the pessimistic meta-induction and the problem of underdetermination have challenged scientific realism and prompted realists to refine their position in various ways. This has led to, e.g., entity realism, perspectivism, and various forms of structuralism---e.g. epistemic à la Worrall \cite{Worrall1989} or ontic à la Ladyman-French \cite{Ladyman1998, French-Landry-Rickles2012}. We also note that although most discussions of GR and gauge symmetries operate within such more or less realist frameworks, there are viable alternatives. The phenomenological tradition, for example, suspends metaphysical commitments and instead takes the experiencing agent, or more generally the subject-object relationship, as its primary field of investigation. It has profoundly influenced several deep thinkers on issues of philosophy of physics: notably Hermann Weyl on spacetime \cite{Ryckman2005}---who famously first articulate the gauge principle \cite{ORaif1997}---and Fritz London in his approach to the quantum measurement problem \cite{French2024}. Given its deep interpretive challenges, the latter has proved to be a particularly fruitful domain for deploying the conceptual resources of phenomenology, including its more ``structuralist" variants; see e.g. \cite{Bitbol1996MQ, Bitbol1998reel} or \cite{Berghofer2023, BerghoferWiltsche, Berghofer-Wiltsche2025}. Here we limit ourselves to drawing attention to the analogy between the interpretive attitudes towards GR and towards GFTs.}
}
the empirical success of GR {is often taken to provide prima facie support for}
a realist view of spacetime described as a 
Lorentzian manifold $(M, g)$, modulo the necessary refinements brought by the hole and point-coincidence arguments.
{By a similar logic,} the empirical success of GFT 
{arguably provides prima facie support for}
 a realist view of the enriched connected spacetime described as a bundle with connection $(P, \omega)$, modulo the necessary refinements brought by generalized hole and point-coincidence arguments.
 
 {It is interesting to remark that while realist/substantivalist positions on space and time predated GR (and SR), the former being significantly updated in light of the latter,\footnote{{With precursors envisioning the geometry of space to be non-Euclidean; e.g. Gauss, Bolyai, Lobachevsky, Riemann, or Clifford \cite{Pesic}. While Bolyai and Lobachevsky merely hinted at the possibility, Riemann and Clifford explicitly argued that the geometry of space was an empirical matter. Gauss famously attempted one such test in his geodetic survey of Hanover kingdom in the  1820s;  in the early 1880s---once Bessel opened in 1838 the era of the measure of star distances via parallax---Zöllner attempted the same type of test using triangles between Earth and two stars.} } 
  no such positions 
 towards an ``enriched" space(time), i.e. a bundle space, existed before the advent of GFT---and that GFT could ``update".
 This may in part explain why bundle substantivalism remains a position seldom articulated and defended.\footnote{{We thank a referee for suggesting to comment on this observation.}}}
 {
 This is arguably on account of the fact that there is no possible ``natural attitude" toward such a space, as it is beyond direct sensory perception/inference; contrary to  space/time, inferred from the immediately perceived spatio-temporal order/quality of phenomena, and thus naturally part of pre-relativistic theoretical frameworks (though Galilean-Newtonian space-time is already somewhat removed from  ordinary, pre-theoretical perception). 
}

{
 Nonetheless, retrospectively,  the development in the 18th and early 19th centuries of \emph{potential theory} (first for the gravitational field, next for the electric fields, then for the magnetic fields) and of Faraday-Maxwell electromagnetic field theory by the mid 19th century,  could arguably be  seen as already weighing in favor of the reality of such enriched space. 
 Indeed, fields (potentials) assign new, non-spatiotemporal qualities, or degrees of freedom, to  points of space. 
 In the case of electromagnetic theory, it was furthermore argued by Faraday and  confirmed by Hertz that these d.o.f have autonomy w.r.t. to those of the ``sources" (electric charges/currents, and magnets), fleshing out an ontological view on (electromagnetic) fields, beyond their former status as mere computational book keeping devices. Decades later, in the late 50s and early 60s, Aharonov and Bohm \cite{AB1959, AB1961} would argue that the effect now named after them, does the same for the electromagnetic potential.
 The new ``qualities" that electromagnetic (EM) field theory introduces, being irreducible to simple kinematic properties `bestowed' by the (Newtonian or special relativistic) spatio-temporal order of phenomena, could have been seen as a genuine `overlay above' space(-time) yielding an ``enriched EM spacetime",   foreshadowing the modern $H=U(1)$ principal bundle picture. 
The EM gauge principle for $\U(1)_{(\text{loc})}$ gauge symmetry, introduced by Weyl in 1929 and fully understood within the modern bundle picture, could then be seen as updating a ``naive" (if not pre-theoretic) view of this enriched spacetime, introducing a relational refinement---like GR did for pre-relativistic and special relativistic spacetime theories.
 A similar logic might have been followed  after the introduction of each new ``quality" (i.e. `internal' d.o.f.) found necessary to account for  the phenomenology (of weak and strong interactions), together with the associated gauge symmetries and bundle pictures.}

 {Our discussion does not rely on one finding  such  reconstructions convincing---though it may help. 
 It is enough to stress that, given the state of the art of established physics, the logic by which one argues in GR for a realist/substantivalist position towards spacetime seen as a Lorentzian manifold, updated by the hole and point-coincidence arguments,  ought to apply \emph{mutatis mutandis} in GFT and to weight in favor of a realist/substantivalist position towards an enriched spacetime seen as a fiber bundles with connection, updated by the generalized hole and point-coincidence arguments.\footnote{{In this we push back against the argument that bundle realism might be primarily motivated by  non-perturbative features of GFT, such as monopoles and instanton solutions, tied to topologically non-trivial bundles. 
 Indeed, as of now, these have little (at best indirect) to no empirical support, contrasting with the considerable wealth of data supporting (perturbative) GFT, i.e. `connection structure on a bundle'. If~that were to change in the future, with non-perturbative topologically non-trivial effects cleanly showing up experimentally, it would `only' be an \emph{additional} (and welcome) support to a bundle realist view. 
 Similarly, pursuing the analogy with GR, if future experiments were to show that spacetime's topology is non-trivial---either at the cosmological scale via correlations in the CMB or via galaxy image pairs in  deep sky surveys, or at the local scale via wormholes detection/creation---it would constitute (admittedly striking) \emph{further} support for ``spacetime as a Lorentz manifold $(M, g)$ 
 realism" piling on the already established immense empirical success of GR. 
We observe that, as far as  we know, it seems that most authors' adhesion  or reservation regarding ``$(M, g)$-realism"  has little to do with spacetime topology.
Let us also stress that the hole and point-coincidence arguments are by essence \emph{local} arguments, they are indifferent to the topology of $M$; the conceptual insight on the nature of spacetime they yield thus holds whatever its topology is---and whether it can be experimentally detected or not. And this is orthogonal to the fact that the very \emph{existence} of some fields is in itself  a constraint on $M$'s topology, notably spinors  (fermions). 
Likewise, as we show next, the internal (and generalized) hole and point-coincidence arguments are local arguments indifferent to the topology of $P$, and thus deliver insights on the nature of the physical enriched spacetime whatever its topology actually is (discovered to be). And this is orthogonal to the fact that the \emph{existence} of some fields is a constraint on $P$'s topology, notably  scalar fields  (Higgs fields)---see \cite{JTF-Ravera2024c} on this. }}
Let us now articulate the case.
 }
\medskip

As reminded in section \ref{Fiber bundles and basics of gauge theory}, GFT physics describes the coupled dynamics of matter and gauge interaction fields. 
The latter are the local, bundle coordinate, representatives of intrinsic objects on a principal bundle $P$. 
So, even if field theories are usually written on $M$,  the fundamental fields are those on $P$: 
The field space is thus $\Phi=\{\,\omega, \alpha\,\}$, with $\alpha$ representing a set of tensorial forms (matter fields, soldering form)---so that if $\alpha =\theta$, then $(\omega, \theta)=\varpi$ is a Cartan connection for a gauge theory of gravity.
The gauge group, or the vertical automorphism group $\H\!\!\simeq\!\!\Aut_v(P)$, partitions $\Phi$ into orbits $\O$.
The moduli space of $(\H\!\!\simeq\!\!\Aut_v(P))$-orbits is the quotient $\M\defeq \Phi/\H$, and there is the projection $\uppi: \Phi \rarrow \M$, s.t. $[\omega, \alpha]=[\omega^\upgamma=\Xi^*\omega, \alpha^\upgamma=\Xi^*\alpha ]$, i.e. there is a 1:1 correspondence $\O \leftrightarrow [\omega, \alpha]$. 

The defining desideratum of GFTs is the $\H_\text{loc}$-covariance of field equations \eqref{cov-E}, whose solutions $\upphi=(A, a)$ are the local representatives of global objects $(\omega, \alpha)$.
The space of global solutions
$\S\defeq\{\, (\omega, \alpha) \in \Phi \ \text{ s.t. }\, E(\upphi)=0\,\}$  is then also partitioned into distinct $\H$-orbits $\O$, so that $\uppi: \S \rarrow \M_\S$, $(\omega, \alpha)\mapsto [\omega, \alpha]$, where $\M_\S$ is the moduli space of global solutions. 
The~immediate dual consequence is that GFT  can neither distinguish solutions in the same  $\H$-orbit $\O$, nor  the points of the fibers, being indifferent to the distinction $p$ \emph{vs} $\Xi(p)=p\upgamma(p)$, both in $ P_{|x} \subset P$.

Here as in GR, 
an a priori realism toward the formalism  amounts to accepting a metaphysical multiplicity of fields, in $\O$, that are physically indistinguishable.
The ``internal" hole argument  further highlights that this position  leads to an ill-defined Cauchy problem: 
Consider {fields on $P$ belonging to the same gauge orbit in solution space,} $(\omega, \alpha)$ and $(\Xi^*\omega, \Xi^*\alpha) \in \O\subset \S$, {s.t. the transformation relating them} $\Xi \in \Aut_v(P)\sim\upgamma \in \H$ has compact support; {the ``internal hole" $P_{|D}$, a `cylinder' in $P$ over a compact region $D \subset U\subset M$, s.t.} $P_{|D}\subset P_{|U} \subset P$.
{Which means that the transformation is non-trivial only on the ``internal hole",}
$\Xi\sim\upgamma\neq \id$ on $P_{|D}$, so that {the gauge-related fields differ only in it but are identical outside of it;} $(\omega, \alpha)=(\omega^\upgamma, \alpha^\upgamma)$ on $P/P_{|D}$, but $(\omega, \alpha)\neq(\omega^\upgamma, \alpha^\upgamma)$ on $P_{|D}$. 
For the local representatives $\upphi=(A, a)$ this means that $\gamma\in \H_\text{loc}$ has compact support on $D$, i.e. $\gamma \neq \id$ on $D\subset M$, so that $\upphi=\upphi^\gamma$ on $M/D$, but $\upphi\neq\upphi^\gamma$ on~$D$.\footnote{Notice that one works here with a given trivializing section $\s:U\rarrow P_{|U}$.
So we are comparing the local representatives on $U$ of \emph{distinct} global objects on $P_{|U}$, fields and their $\H$-transforms, through a unique local section. 
It is different from comparing local representatives on $U$ of the \emph{same} global objects, through distinct local sections $\s$ and $\s'$: this is the operation of ``gluings" via transition functions of $P$, i.e. of bundle coordinate change, and represent \emph{passive gauge transformations} akin to coordinate change in GR. These do not lead to ontological underdetermination, nor to indeterminism, as there is no hole argument holding. {This is the internal counterpart of Stachel's key remark \cite{Stachel1986, EarmanNorton, Stachel2014} that the (external) hole argument has force only with ``active" diffeomorphisms $\Diff(M)$, not with ``passive" ones, i.e. mere coordinate changes. See \cite{JTF-Ravera2024c} for a precise articulation of this point. \label{Stachel-active-diff}} } 
Then, given boundary conditions for the fields  $\upphi$ at $\d U$---or for $(\omega,\alpha)$ at $P_{|\d U}$---the field equations {cannot} uniquely determine  a solution within $U$, as gauge related ones like $\upphi$ and $\upphi^\gamma$ can differ on $D \subset U${, yet share the same boundary conditions}. 
{Thus, } 
a global solution in $P_{|U}$ { cannot be determined either,} since both $(\omega, \alpha)$ and $(\Xi^*\omega, \Xi^*\alpha) \in \O\subset \S$. 
This in particular implies a failure of determinism if the domain $D$ is in the future of some Cauchy surface for the fields. 

The natural way out, as in GR, is to appeal to an ``internal point-coincidence argument", which emphasizes that the physical content of a GFT (in particular, its 
empirically accessible predictions) is exhausted by the pointwise coincidental values of fields' internal d.o.f.---{meaning the relation
``[value of $\omega$ at $p$] $\leftrightarrow$ [value of $\alpha$ at $p$]"---which are $\Aut_v(P)\!\simeq\!\H$-invariant,}
{and thus \emph{define} what a ``physical internal point/event" \emph{is}.}
The immediate dual consequences are as follows.

First, points $p\in P_{|x}$ of the fibers, the internal space over/within $x=\pi(p) \in M$, are entirely unphysical. 
{\emph{Physical fiber spaces}, i.e. all the physical internal points, are (in \emph{actuality}) realized as} 
the $\Aut_v(P)$-invariant 
network of relations among the internal d.o.f. of the fields $(\omega, \alpha)$/$\upphi$; 
{Together, the internal hole and point-coincidence arguments thus paint a relationalist view of the \emph{physical internal space} of GFT, which is indeed an ``anti-haecceitism for physical internal points".}
Secondly, the argument implies the standard physics view and practice, which is to consider
that \emph{physical fields} d.o.f. are not represented by any single $\upphi$, or $(\omega, \alpha) \in \O\subset \S $, but by a point in a $\H_\text{(loc)}$-equivalence class $[\upphi]$, or $[\omega, \alpha] \in \M_\S$ in the moduli space of solutions. 
This extends Earman and Norton's ``Leibniz equivalence"~to~GFT.
The~argument further suggests that what is physical are the $\Aut_v(P)\simeq\H$-invariant relations among the internal d.o.f. of the fields $\upphi$/$(\omega, \alpha)$, so that physical fields internal d.o.f. mutually co-define each other: e.g. the internal d.o.f. of the gauge fields (redundantly described by $A/\omega$) are only well-defined in relation to those of the matter fields (redundantly described by $\vphi/\phi$). 
\medskip

In view of equation \eqref{SES-P}, the standard and internal hole arguments combine into a \emph{generalized hole argument} which, 
articulated with a \emph{generalized point-coincidence argument}, 
makes a strong case for the relationalist view of the principal bundle: 
The fibered manifold $P$ only bootstraps our ability to erect a description of the relevant physics of fields with spatio-temporal and internal d.o.f.,\footnote{
In the same way that in GR, \cite{Einstein1952-EINRTS-3} stressed that fields are not ``in spacetime", but have spacetime extension i.e. spatio-temporal d.o.f., 
in GFT one may say that the fields do not in part ``live" or ``probe" an independently existing internal space at each spatio-temporal event $x$---the fibers $P_{|x} \subset P$---but have an ``internal extension", i.e. internal d.o.f. 
}
but is removed from the physical picture by the requirement \eqref{cov-E} of covariance of the field equations   under $\Diff(M)\ltimes \H_\text{loc}$, {which implies (or can be understood to stem from) sensitivity of  gRGFTs to  $\Aut(P)$-orbits of global objects only.}
To paraphrase \cite{Einstein1952-EINRTS-3}, the \emph{physical fibered manifold of spatio-temporal and internal events}  does not claim existence on its own, but only as a \emph{structural quality of the fields}{---represented by $(\omega, \alpha)$}. 

One may define a notion of ``enriched spacetime" as being represented by $(P, \omega)$, understood that $\omega \!=\! \omega'\!\oplus~\!\varpi$, with~$\omega'$ an Ehresmann connection describing Yang-Mills gauge interactions fields, and $\varpi$ a Cartan connection describing the gravitational field. 
A physical gauge field, represented by $\omega$, is as autonomous an entity as is possible in GFT, with the
physical field d.o.f. and the co-defining relations in which they participate being coextensive. 
Then one may consider physical gauge fields to embody a substantive notion of enriched spacetime, insofar as the “substantival”
\enlargethispage{1\baselineskip}
notion gets suitably “relationalized”.
This would be a form of ``sophisticated substantivalism" towards enriched spacetime, {exactly} analogous to sophisticated substantivalism towards spacetime.\footnote{ \cite{Healey2001} pointed out that, analogously to metric essentialism, one could define ``connection essentialism"  
(though he rejects it).}

It can be understood within a generalization of \cite{Norton1989}'s ``mpfs", which we may call ``enriched manifolds plus further structures", ``empfs". The further structure above being the connection $\omega$. 
Emulating  \cite{Norton1989}'s argument, one could argue that empfs substantivalism is vulnerable to a hole-type argument if the ``further structure" has a non-trivial Killing  symmetry group. 
But we could first 
respond that realistic fields, even in best cases, have only approximate symmetries, 
and secondly that a  \cite{Norton1989} type argument cannot hold in  a fully \emph{dynamically coupled regime}.\footnote{Where e.g. no Killing gauge symmetries of a Yang-Mills gauge potential  can fail to be symmetries of the matter fields due to the Yang-Mills equation $E(A, \vphi)= D *\!F - J(\vphi, A)=0$, where $J(\vphi, A)$ is the current $(m-1)$-form of the matter field, and $*F \in \Omega^{m-2}(U, \LieH)$ is the Hodge dual of the curvature/field strength of $A$.}

The significance of the covariance \eqref{cov-E} of the field equation of gRGFT under $\Diff(M)\ltimes \H_\text{loc}$, analyzed in the framework's own terms,
is decisively clarified by the generalized hole and point-coincidence arguments, yielding a relationalist view of $P$ and a correspondingly relationalized  sophisticated substantivalism towards the field of gRGFT. 
In the next section, we describe how the \emph{Dressing Field Method} (DFM) allows a technical framing of these insights, through the definition of invariant dressed fields and dressed spaces.

\section{Dressing Field Method: Dressed fields and spaces}\label{Dressing Field Method: Dressed fields and spaces}

As reflected in the above discussion, it is the established view in physics that 
physical magnitudes and observables must be invariant.
This requirement is typically addressed by appropriately reducing the {local} symmetries of the theory.
The \emph{Dressing Field Method} (DFM) 
is an approach allowing to do so systematically by producing invariant variables out of the field space $\Phi$ of a theory. 
{It is a unified framework encompassing many constructions found in the physics literature, old and new,    streamlining their technical underpinning and clarifying the conceptual landscape: e.g., Stueckelberg fields \cite{Ruegg-Ruiz} are instances of \emph{ad hoc} dressing fields, so are
 `edge modes' à la \cite{DonnellyFreidel2016}.
Special cases of the DFM notably include: Dirac variables \cite{Dirac55, Dirac58}, asymptotic dressed states à la Faddeev-Kulish \cite{Kulish-Faddeev1970, Akhoury-et-al2013, Gabai-Sever2016}, `composite operators' à la Fröhlich-Morchio-Strocchi (FMS) \cite{Frohlich-Morchio-Strocchi80, Maas2019}, Bardeen-Wess-Zumino anomaly cancelling counter-terms \cite{Manes-Stora-Zumino1985, Bertlmann, Bonora2023}, `gravitational dressings' à la \cite{Giddings-et-al2006, Giddings-Perkins2024},  `dressing 2-form' à la \cite{De-Paoli-Speziale2018, Oliveri-Speziale2020}, `dynamical reference frames' à la \cite{Carrozza-Hoehn2021, Hoehn-et-al2022}, etc.}
See 
\cite{JTF-Ravera2024gRGFT} (and references therein) for the most up-to-date technical account, \cite{JTF-Ravera2025DFMSusyReview&Misu, JTF-Ravera2024RaritaSchwinger, JTF-Ravera2025UnconventionalSusy} for recent applications to  supersymmetric field theory and supergravity,  \cite{JTF-Ravera2024NRrelQM, JTF-Ravera-MFT2025} for applications to quantum mechanics and anomaly cancellation. 

What makes the DFM particularly attractive from a philosophical perspective is that 
by allowing to technically implement a manifestly invariant reformulation of 
gauge theories, it ``wears its ontology on its sleeve", 
so to speak. 
Hence~it gained some traction in the foundations of 
gauge theories, see e.g. \cite{Berghofer-et-al2023, Berghofer-Francois2024, JTF-Ravera2024c}.

In what follows, we remind the basics of the DFM, introducing dressing fields, dressed fields, and dressed spaces. 
We shall see how the DFM technically implements the point-coincidence argument, meaning  that dressed spaces are by definition immune to the hole argument, thus providing a transparent formal implementation of the conceptual insights discussed in sections \ref{The hole argument} and \ref{The generalized hole argument}.

\subsection{The case of internal gauge symmetries}
\label{The case of internal gauge symmetries}

Consider a GFT with field content $\upphi=\lbrace{ A, \varphi \rbrace}$ and (internal) gauge group $\H_\textbf{loc}$. 
A \emph{dressing field} is a smooth map
\begin{align}
\label{dressing-field-YM}
    u : U\subset M \rightarrow H , \quad \text{defined by } \quad u^\gamma = \gamma\- u ,
    \quad \gamma \in \H_\textbf{loc} .
\end{align}
Key to the DFM is the fact that a dressing field should be extracted from the d.o.f. in the field space $\Phi$ of the theory. 
This means that it should be a \emph{field-dependent dressing field}, $u=u[\upphi]$, so that its defining gauge transformation is
\begin{align}
\label{dressing-field-YM-field-dep}
    u[\upphi]^\gamma := u[\upphi^\gamma] = \gamma\- u [\upphi] , \quad \gamma \in \H_\textbf{loc} .
\end{align}
Given  a dressing field as above, one defines the $\H_\text{loc}$-invariant \emph{dressed fields}, as ``composite objects", as follows:
\begin{align}
\label{dressed-fields}
    \upphi^u 
    = \big\{\, A^u , \, \varphi^u \,\big\} 
    \defeq
    \big\{ \, u\- A u + u\- du, \ \  \rho(u)\- \vphi \,\big\}.
\end{align}
To anticipate a contrast with the forthcoming discussion of dressings for $\Diff(M)$, notice that $\upphi^u$ live on $M$ still.
This is the simplest illustration of the DFM ``rule of thumb": To dress any object (fields, functional of fields), first compute  its gauge transformation, then formally substitute in the resulting expression the gauge parameter $\gamma$ by the dressing field $u$. 
The new expression is the ``dressing" of the object and is by construction $\H_\text{loc}$-invariant. 

It must be stressed that since the dressing field \emph{is not} an element of the gauge group, as it does not have the defining $\H_\text{loc}$-transformation of such an element---see eq. \eqref{loc-gaugr-grp}---the dressed fields \eqref{dressed-fields} are therefore not gauge-transformed fields, and, in particular, 
a dressing via the DFM \emph{is not} a gauge fixing. 
See \cite{Berghofer-Francois2024,JTF-Ravera2025DFMSusyReview&Misu} for a detailed discussion of this point.

Relatedly, notice that with 
 $\upphi$-dependent dressing fields, the DFM has a very natural \emph{relational} interpretation.
The~dressed fields
$\upphi^{u[\upphi
]}= \{A^{u[A, \vphi]}, \vphi^{u[A, \vphi]}\}$
can be understood as resulting from a reshuffling of the d.o.f. of the ``bare" fields $\upphi$, which mixes physical d.o.f. and non-physical pure gauge modes, whereby  pure gauge modes are eliminated. 
Said otherwise, some of the internal  d.o.f. of the bare fields $\upphi$ are used (in $u[\upphi]$) to  ``coordinatize" 
the remaining internal d.o.f. of $\upphi$ (yielding $\upphi^{u[\upphi]}$),  invariantly~so.
The dressed fields \eqref{dressed-fields} thus 
represent the \emph{gauge-invariant} \emph{relations} co-defining the physical internal d.o.f. embedded in  bare fields $\upphi=\{A, \vphi\}$, 
they are ``relational variables". 
Furthermore, insofar as   dressed fields $\upphi^{u[\upphi]}$ achieve gauge invariance  relationally---whereby physical internal field d.o.f. coordinatize, even co-define, each other---they  formally implement the 
\emph{internal point-coincidence argument} as discussed in section \ref{The generalized hole argument}, and are thus immune to the \emph{internal hole argument}.
See \cite{JTF-Ravera2024gRGFT} for   further discussion of this point, illustrating it in the simple case of scalar ($\CC$-) electrodynamics (section 5.2.7). 
Notably, all this is achieved already at the kinematical level.
\medskip

The dynamics of the theory being specified by a  Lagrangian $L(\upphi)=L(A, \varphi)$  that is $\H_\textbf{loc}$-quasi-invariant, eq. \eqref{Sym-L}, 
given  a dressing field $u$, by the DFM rule of thumb one  defines the~\emph{dressed~Lagrangian}
\begin{equation}
\label{dressed-Lagrangian}
L^u \defeq L(\upphi^u)=L(\upphi) + db(u;\upphi),
\end{equation}
which is just the Lagrangian expressed in terms of the dressed fields. 
If $L$ is strictly $\H_\textbf{loc}$-invariant, i.e.  $b=0$, then $L(\upphi^u)=L(\upphi)$. 
In either cases, the field equations $\bs E(\upphi^u)=0$ for the dressed fields, obtained from \eqref{dressed-Lagrangian},  have the \emph{same functional expression} as the field equations $\bs E(\upphi)=0$ for the bare fields, obtained from $L$. 
Manifestly, here they only differ by a boundary term: i.e. $\bs E(\upphi^u)=\bs E(\upphi)+ dc(u; \upphi)$, see eq. (391) in \cite{JTF-Ravera2024gRGFT}.
The~dressed field equations are \emph{deterministic}, in the sense that they uniquely determine the evolution of the relational d.o.f. $\upphi^u$ of the theory, as expected from a scheme that solves the problem posed by the internal hole argument by technically implementing the internal point-coincidence argument in a manifest way.

{It is worth noticing that, in the spirit of the internal point-coincidence argument, it is  neither the bare nor a gauge-fixed version of a theory that is confronted to experiment, but  
a version  expressed via invariant, relational variables; i.e. a dressed version like \eqref{dressed-Lagrangian}, in the present framework.  
}

\subsection{The case of diffeomorphisms}
\label{The case of diffeomorphisms}

We now consider a general-relativistic theory, with field content $\upphi=\big\{A, \vphi, e/g \big\}$ subject to the action of $\Diff(M)$.
A~$\Diff(M)$-dressing field, introduced in
\cite{Francois2023-a}, is a smooth map
\begin{align}
\label{dressing-field-diffeo}
    \upsilon: N \rightarrow M, \quad \text{defined by} \quad \upsilon^\psi := \psi\- \circ \upsilon,
    \quad 
    \psi \in \Diff(M).
\end{align}
Again, key to the DFM is the notion that the dressing field should be extracted from the d.o.f. of the theory, i.e. it should be a $\upphi$-dependent dressing field $\upsilon = \upsilon[\upphi]$, so that its $\Diff(M)$-transformation is  
\begin{align}
\label{dressing-field-diffeo-field-dep}
    \upsilon[\upphi]^\psi := \upsilon [\psi^* \upphi] = \psi\- \circ \upsilon[\upphi].
\end{align}
Given a dressing field $\upsilon$ as above, we define the $\Diff(M)$-invariant dressed fields
\begin{equation}
\label{diffeo-dressed-fields}
\begin{aligned}
     \upphi^\upsilon := \upsilon^* \upphi, 
     \quad 
    \text{i.e.} \quad 
    \lbrace{ A^\upsilon, \varphi^\upsilon, e^\upsilon/g^\upsilon \rbrace} = \lbrace{ \upsilon^* A, \upsilon^* \varphi, \upsilon^*e/\upsilon^* g \rbrace} .
\end{aligned}
\end{equation}
Again, this is a simple case of the DFM rule of thumb:
To dress any object---field or functional thereof---one writes its $\Diff(M)$-transformation and then formally replaces $\psi$  in the resulting expression with the dressing field $\upsilon$. 
The~new expression is the dressing of the object, and is $\Diff(M)$-invariant by construction.

The dressed fields $\upphi^{\upsilon[\upphi]}$ again admit a natural relational interpretation.
Some of the spatio-temporal  d.o.f. of the bare fields $\upphi$ are used (in $\upsilon[\upphi]$) to  invariantly coordinatize 
the remaining spatio-temporal d.o.f. of $\upphi$, thereby yielding $\upphi^{\upsilon[\upphi]}$.
The dressed fields then 
represent the $\Diff(M)$-\emph{invariant}  \emph{relations} co-defining the physical d.o.f. embedded in  bare fields $\upphi$;
they are relational variables indeed.
Seeing that they achieve $\Diff(M)$-invariance relationally, with physical spatiotemporal field d.o.f.
co-defining each other as coincidental values of fields,
the dressed fields \eqref{diffeo-dressed-fields} formally implement the \emph{point-coincidence argument} as discussed in section \ref{The hole argument}, and are therefore immune to the \emph{hole argument}. 
All this is achieved at the kinematical level.
Again, see \cite{JTF-Ravera2024gRGFT} for a thorough discussion, illustrating the point in the case of GR with a set of scalars (section 5.2.7)---on this see also the end of this section.

The dynamics of the theory being specified by a Lagrangian $L(\upphi)$ transforming under $\Diff(M)$ as in eq. \eqref{Sym-L}, given a dressing field $\upsilon$  one defines the dressed Lagrangian as 
\begin{equation}
\label{diff-dressed-L}
L^\upsilon\defeq  L(\upphi^\upsilon)=\upsilon^* L(\upphi),
\end{equation}
which is strictly $\Diff(M)$-invariant by construction. 
This is yet another application of the DFM rule of thumb. 
The~field equations $E(\upphi^\upsilon)=0$ for the dressed fields have the same functional expression as those $E(\upphi)=0$ for the bare fields: they are related by $E(\upphi^\upsilon) = \upsilon^* \big(E(\upphi) + dc(\upsilon; \upphi) \big)$, see eq. (391) in \cite{JTF-Ravera2024gRGFT}. 
But~contrary to the bare ones, the dressed field equations are deterministic. 
{Here again we may observe that, in the spirit of the point-coincidence argument, what is confronted to empirical tests is a version of a theory expressed via invariant relational variables, such as the  dressed version  \eqref{diff-dressed-L} indeed, rather than the bare (or a gauge-fixed) version.}
\medskip

Contrary to the case discussed in section \ref{The case of internal gauge symmetries}, 
observe that the dressed fields $\upphi^\upsilon$ do \emph{not} live on ``bare regions" $U\subset M$, but on  \emph{field-dependent dressed regions} defined as
\begin{align}
\label{dressed-regions}
    U^\upsilon := \upsilon\- (U) = \upsilon[\upphi]\- (U),
\end{align}
where $\upsilon\-$ is the inverse map of $\upsilon$, s.t. $\upsilon \circ \upsilon\-=\id_M$. 
These dressed regions are $\Diff(M)$-invariant: 
\begin{align}
    (U^\upsilon)^\psi = (\upsilon[\upphi]^\psi)\- \circ (U^\psi) = \upsilon[\upphi]\- \circ \psi \circ \psi\- \circ (U) = U^\upsilon.
\end{align}
The $\upphi$-dependent $\Diff(M)$-invariant regions $U^{\upsilon[\upphi]}$ represent \emph{physical regions} of the physical manifold of events, and are thus obviously immune to the hole argument. 
The definition \eqref{dressed-regions} comes from integration theory, telling that integration is a $\Diff(M)$-invariant operation; e.g.  $S\defeq\!\int_U L(\upphi) =\!\int_{\,\psi\-(U)}\psi^*L(\upphi)\,$  yields by the DFM rule of thumb,
\begin{equation}
\label{int-dressed-fields}
S=\int_U L(\upphi) =\int_{\,\upsilon\-(U)} \upsilon^*L(\upphi)\rdefeq S^\upsilon.
\end{equation}  
In particular, a physical spatio-temporal event is a $\upphi$-dependent $\Diff(M)$-invariant point $x^{\upsilon[\upphi]}\defeq \upsilon[\upphi]\-(x) \in U^\upsilon$, 
which~is the technical implementation of the basic intuition underlying the point-coincidence argument. 
Remark that the latter was always partly tacitly baked into integration theory: 
Indeed, integration of a 0-form $\vphi$ over the 0-dimensional manifold $x \subset M$  is the $\Diff(M)$-invariant evaluation operation ${ev}_x (\upphi) \defeq \int_x \vphi = \int_{\,\psi\-(x)} \psi^* \vphi$ giving the invariant ``value of $\vphi$ at the point $x$", which can be  understood as saying that the ``value of $\vphi$" and ``the point $x$" are co-defining, or relata coextensive with the relation they define.  
The DFM rule of thumb thus yields
\begin{equation}
\label{eval-dressed-fields}
  {ev}_x (\upphi) = \int_x \vphi = \int_{\upsilon\-(x)} \upsilon^* \vphi \rdefeq {ev}_{x^\upsilon} (\upphi^\upsilon).
\end{equation} 
Equations \eqref{int-dressed-fields}-\eqref{eval-dressed-fields} implement in a mathematically concrete way the conceptual insight, described in section \ref{The hole argument}, according to which relational field d.o.f. ``live" on the relationally defined physical regions. 
See \cite{JTF-Ravera2024gRGFT} for technical details, which involve the definition of integration as an operation on the \emph{bundle of regions} (of $M$) associated to the field space bundle $\Phi$. 

One may then define 
$M^\upsilon \defeq \im(\upsilon\-)$, which is identified as the \emph{physical manifold of spatio-temporal events}. 
Notice that $M^\upsilon$, its regions $U^\upsilon$ and points $x^\upsilon$,  
simply \emph{do not exist} without the fields $\upphi$, on which they are \emph{supervenient},  and are thus the precise formal encapsulation of \cite{Einstein1952-EINRTS-3}'s remark, quoted already at the end of section \ref{The hole argument}, saying  
``\emph{If we imagine the [gravitational and other]
field[s] [...] to be removed, there does not remain a space of the type (1) [of SR], but
\textbf{absolutely nothing}, and also no `topological space'. [...]
Space-time does not claim existence on its
own, but only as a structural quality of the field}" (our emphasis).
While this is \emph{tacitly} encoded by the $\Diff(M)$-covariance of general relativistic-theories, 
it is made \emph{manifest} via the DFM. 

The \emph{physical spacetime} may then be defined as the pair $(M^\upsilon,g^\upsilon)$, which 
formally  realizes the ``relationalized" sophisticated substantivalism emphasized in section \ref{The hole argument}. 
Alternatively, defining spacetime as $(M^\upsilon, {\mathscr S}^\upsilon)$, for ${\mathscr S}$ any other structure besides, or instead of, a metric---most likely an Ehresmann (spin) connection  ${\mathscr S}=A$, or   ${\mathscr S}=(A, e)$~a~Cartan connection---we obtain a ``relationalized" sophisticated substantivalism of the ``mpfs" kind. 
Clearly, Norton's worry \cite{Norton1989} that ``mpfs substantivalism" might be vulnerable to hole-type arguments cannot even get off the ground here.

The above is the general framework  encompassing all manners of ``scalar coordinatization" in GR: 
e.g. the Kretschmann-Komar  approach  \cite{Komar1958} where $\upsilon=\upsilon[g]$ and $g^{\upsilon[g]}$ is a case of self-dressing, or approaches à la {DeWitt \cite{DeWitt1960}  or} Brown-Kucha\v{r} \cite{Brown-Kuchar1995} where $\upsilon=\upsilon[\vphi]$ and $\vphi$ is some effective dust scalar field so that $g^{\upsilon[\vphi]}$ is the metric as seen in the ``coordinate system" provided by the dust field, which is also a point of view developed by \cite{Rovelli1991, Rovelli2002b}, and one that has been applied e.g. to cosmology by \cite{Giesel-et-al2010}.

\subsection{The case of general-relativistic gauge field theories}
\label{The case of general-relativistic gauge field theories}

Bringing together the results of sections \ref{The case of internal gauge symmetries} and \ref{The case of diffeomorphisms}, the DFM allows to reformulate gRGFTs in a manifestly invariant way:
Defining a complete dressing field as the pair $(\upsilon,u)$ satisfying  \eqref{dressing-field-YM} and \eqref{dressing-field-diffeo},  the dressed fields 
\begin{align}
\label{dressed-fields-gRGFT0}
\upphi^{(\upsilon, u)}
\defeq    
\upsilon^*(\upphi^u)
\end{align}
are $\big(\!\Diff(M)\ltimes \H_\text{loc}\big)$-invariant by construction. 
These are fully relational variables, representing invariant relations among the d.o.f. of the bare fields $\upphi$, formally implementing the \emph{generalized point-coincidence} argument and thus immune to the \emph{generalized hole argument}, as discussed in section \ref{The generalized hole argument}.
They represent \emph{physical} spatio-temporal and internal d.o.f. of fields whose dynamics is given by the $\big(\!\Diff(M)\ltimes \H_\text{loc}\big)$-invariant dressed Lagrangian 
\begin{equation}
    L^{(\upsilon, u)}
\defeq    
L(\upsilon^*(\upphi^u))
= \upsilon^* L((\upphi^u)),
\end{equation}
yielding the fully deterministic (dressed) field equations 
$E(\upphi^{(\upsilon, u)})=0$.
Again, these are functionally the same as those of the bare fields---to which they relate by eq. (390) in \cite{JTF-Ravera2024gRGFT}---which explains why gRGFTs could be  empirically tested \emph{before} the issue of their observables could be cleanly settled: it is the dressed theories that were put to test. 
A fact e.g. stressed, in his own language, by \cite{Rovelli2002b} in the case of GR. 

As previously, the fields $\upphi^{(\upsilon, u)}$ live on the physical manifold of spatio-temporal events $M^{(\upsilon, u)}= M^\upsilon$ (or regions $U^\upsilon$ thereof),  immune to the hole argument. 
One may rightly suspect that the above is a \emph{local} description of a global intrinsic structure: that of global dressed fields on a dressed bundle space. 

Indeed, in the same way that $(\psi, \gamma) \in \Diff(M)\ltimes \H_\text{loc}$ is the local (coordinate bundle) version of a global automorphism $\Xi \in \Aut(P)$, the pair 
$(\upsilon, u)$ is the local version of a global dressing field, which is a smooth map
\begin{align}
\label{dressing-field-diffeo-new}
    \u: Q \rightarrow P, \quad \text{defined by} \quad \u^\Xi := \Xi\- \circ \u,
    \quad 
    \Xi \in \Aut(P),
\end{align}
where $Q$ is a ``model" fibered manifold. Remark that $\u$ is not a \emph{principal} bundle morphism. 
Such a dressing field is built from the  d.o.f. of the global fields $\b\upphi=(\omega, \alpha)$ on $P$---with local representatives on $U \subset M$ the bare fields~$\upphi$. 
It~is a $\b\upphi$-dependent dressing field $\u = \u[\b\upphi]$, so that its defining $\Aut(P)$-transformation is  
\begin{align}
    \u[\b\upphi]^\Xi := \upsilon [\Xi^* \b\upphi] = \Xi\- \circ \u[\b\upphi].
\end{align}
It allows to define $\Aut(P)$-invariant dressed fields 
\begin{equation}
\label{dressed-fields-gRGFT}
    \b\upphi^\u \defeq\u^*\b\upphi,
\end{equation}
whose local representatives are \eqref{dressed-fields-gRGFT0}, and like them are immune to the generalized hole argument. 
They live on the \emph{physical manifold of spatio-temporal and internal events}, or \emph{physical bundle space}, defined as $P^{\,\u}\defeq\im(\u\-)$. 
Its~regions are 
\begin{equation}
\label{dressed-regions-gRGFT}
V^\u\defeq \u\-(V)
\end{equation}
for $V\subset P$, s.t. $\pi(V)=U \subset M$,  and are $\Aut(P)$-invariant:
$(V^\u)^\Xi = \Big(\u[\b\upphi]^\Xi \Big)\- \!\!\circ \big(\Xi \circ (V)\big) = \u[\b\upphi]\- \circ \Xi\- \circ \Xi \circ (V) = V^\u$.
Both dressed fields \eqref{dressed-fields-gRGFT} and regions \eqref{dressed-regions-gRGFT} are thus immune to the generalized hole argument. 
The reason for the definition \eqref{dressed-regions-gRGFT} being, as previously, integration theory (on P). It gives e.g. formal substance to the generalized point-coincidence argument via the notion of the $\Aut(P)$-invariant $\b\upphi$-dependent \emph{physical} point $p^{u[\b\upphi]}\defeq \u[\b\upphi]\-(p)$, which is both a spatio-temporal and an internal physical event.
Indeed, as a special case of \eqref{dressed-regions-gRGFT} we have that the \emph{physical internal space}, the dressed fibers, are $P^{\,\u}_{|x^{\,\u}}\defeq \u\-(P_{|x}) \subset P^{\,\u}$, realizing the internal point-coincidence argument.
This~means that $P^{\,\u}$ is a fibered manifold: the moduli space of physical fibers forming the physical manifold of spatio-temporal events $M^{\,\u}=M^\upsilon$, there is a smooth projection $\b\pi: P^{\,\u} \rarrow M^\upsilon$, $p^{\,\u} \mapsto \b\pi(p^{\,\u})\rdefeq x^{\,\u}=x^\upsilon$. Observe that the physical bundle $P^{\,\u}$---its regions, fibers and points---does not exist without the fields $\b\upphi$ on which it supervenes. 
It formally realizes our paraphrasing of Einstein in section \ref{The generalized hole argument}: the physical fibered
manifold of spatio-temporal and internal events  does not claim existence on its own, but only as a structural
quality of the fields. 

The physical enriched spacetime may then be defined as the pair $(P^{\,\u}, \omega^\u )$, where  $\omega^\u={\omega'}^\u \oplus \varpi^\u$ for $\omega$~and~$\varpi$ Ehresmann and Cartan connections with local representatives $A'$ and $(A,e)$ respectively. 
It ``projects" on the \mbox{physical} spacetime $\big(M^\upsilon, e^{(\upsilon, u)}/g^\upsilon, A^{(\upsilon, u)}, {A'}^{(\upsilon, u)}\big)$, where field theory is usually written. 
This enriched spacetime thus realizes the ``relationalized sophisticated substantival" view. 

\subsection{{Ambiguity in the choice of dressing and physical frame covariance}}
\label{Transformations of the second kind}

The DFM would allow for a particularly clear ontology if the choice of dressing were unique, or unique up to a finite residual freedom. 
If so, it would provide what is called in  \cite{Jacobs2022}  a ``perspicuous" or ``intrinsic formalism" for gRGFTs (see \cite{March2025} for a further discussion of the value of intrinsic formalisms). 
Of course, in \cite{Jacobs2022} the author is primarily concerned with internal sophistication and external sophistication, arguing that only the former can succeed in offering an intrinsic formalism. 
Regarding the distinction between internal sophistication, external sophistication, and reduction (see \cite{Dewar2019,Jacobs2022}): The DFM can be understood as a form of \emph{reduction}, whose aim is characterized in \cite{Jacobs2022} as ``to construct a theory whose models uniquely correspond to equivalence classes of [symmetry-related models] of the old theory. The standard way to achieve this is to reformulate the theory solely in terms of invariant quantities [...]". 
Yet, the paper remarks, reduction can be
``brute forced" through simply quotienting by the relevant covariance group, i.e. by considering equivalent classes (which in but rare cases is impractical). The DFM may then be a ``non-brute force" reduction, a form of `sophisticated reduction' perhaps.  
This goes to show that it does not neatly fit the existing nomenclatures.

The DFM does not, \emph{a priori}, claim that there should be a unique choice of dressing field. This is bound to be a model-dependent issue, although it is plausible that the more sophisticated and realistic the model, the more constrained the available choices become.
Still, the DFM naturally accounts technically for possible ambiguities in the choice of dressing field through what have been called ``transformations of the second kind" (\cite{JTF-Ravera2024gRGFT}).
In what follows, we describe how they arise in local gRGFT. We first consider the case of GFT to illustrate the discussion and then turn to the question how they should be understood.
\medskip

Given their general definition \eqref{dressing-field-YM}-\eqref{dressing-field-YM-field-dep}, 
two $\H_\text{loc}$ dressing fields $u, u'$ may a priori be related by 
$u'=u\xi$, with $\xi \in \G\defeq \{ \xi: U\rarrow G\, |\,  \xi^\gamma=\xi\}$, for $G\supseteq H$, and $\G$ 
is a group under pointwise product.
We may write its action on a dressing field as $u^\xi\defeq u\xi$, and its action on bare variables as $\upphi^\xi$ , so that it transforms dressed fields by $(\upphi^u)^\xi:= (\upphi^\xi)^{u\xi}$. 
In the noteworthy special case
where $\G$ leaves $\upphi$ invariant, one has  $(\upphi^u)^\xi:= (\upphi)^{u\xi}$, meaning that  the $\G$-transformation of the dressed fields is formally exactly the same as the $\H_\text{loc}$-transformation of the bare fields. In this case, the dressed theory is automatically $\G$-(quasi-)invariant, for the same (formal) reason the bare theory was $\H_\text{loc}$-(quasi-)invariant. 
In the generic case, where $\upphi^\xi\neq \upphi$, the form of $(\upphi^u)^\xi$ must be worked out case by case. 
Remark that, at this stage, although $\G$ looks like a new gauge symmetry, somewhat isomorphic to $\H_\text{loc}$, it is generically a distinct mathematical object: a group of gauge-invariant maps. 
The action $(\upphi^u)^\xi$ thus appears to a priori generate a whole $\G$-orbit, that is, a family of $\H_\text{loc}$-invariant objects. 
The relevance of this observation should, however, be assessed in light of the key desideratum that 
$u$ should be a $\upphi$-dependent dressing field $u=u[\upphi]$, see eq. \eqref{dressing-field-YM-field-dep}. 
Indeed, $\G$ will typically emerge as a byproduct of the constructive procedure by which $u$ is built from the existing fields.
This gives rise to two quite distinct situations, formally and conceptually: $\upphi$-dependent and independent variants of $\G$. 

First, in many cases (possibly all) the procedure may rely on some functional constraint $\C(\upphi^u)=0$ that one solves explicitly (if only formally) for $u$ in terms of $\upphi$. Having done so, one then finds
the solution $u=u[\upphi]$ to satisfy the defining dressing property \eqref{dressing-field-YM-field-dep}.\footnote{Or one could enforce it from the outset, by requiring that the constraint be $\H_\text{loc}$-invariant, $\C\left((\upphi^\u)^\gamma\right)=\C(\upphi^\u)$; it is clear that the solution for $u$ must then possess the dressing property \eqref{dressing-field-YM}-\eqref{dressing-field-YM-field-dep}.}
Let us denote by $\G_{[\varnothing]}$ the subgroup of $\upphi$-\emph{independent} elements of $\G$. These may appear either a) as the `homogeneous' $\upphi$-independent solutions of the constraint (e.g. if it is some PDE), or b) as reflecting the inevitable formal freedom to insert the identity operation somewhere in the construction of $u$, which very rarely (as far as we know) will be mathematically well-motivated, thus of little physical and conceptual relevance. 
Nevertheless, in either case one obtains a $\G_{[\varnothing]}$-orbit, or family, of $\H_\text{loc}$-invariant relational dressed variables.  We shall see examples below. 

Then, in a given theory there may be distinct fields' d.o.f. that can serve as candidate dressing fields. Given the fields of the theory, $\upphi=\{\upphi_1, \upphi_2\}$, it may be that one has $u[\upphi_1]$, say built from $\C\left((\upphi_1)^u\right)=0$, and $u'[\upphi_2]$, built from $\C'\big((\upphi_2)^{u'}\big)=0$. 
Since both satisfy the defining property \eqref{dressing-field-YM-field-dep}, they must be related by 
a $\upphi$-\emph{dependent} $\G$ via 
$u'= u\, \xi[\upphi]$; or, seen another way, they  \emph{define} it by $\xi[\upphi]\defeq u[\upphi_1]\- u'[\upphi_2]$.
Let us denote by $\G_{[\upphi]}$ the subgroup of such $\upphi$-dependent elements of $\G$. 
Since $\upphi$-dependent dressing fields $u[\upphi]$ are naturally understood as (internal) \emph{physical reference frames}, with some d.o.f. coordinatizing others, 
the action of $\G_{[\upphi]}$ on dressed fields by $\upphi^u \mapsto (\upphi^u)^{\xi[\upphi]}$ is naturally interpreted as a change of \emph{physical reference frame}. 
Furthermore, by the definition of $\xi[\upphi]$ it seems clear that $\upphi^{\xi[\upphi]}=\upphi$. 
So~that the $\G_{[\upphi]}$-transformations of $\upphi^u$ formally mimic the $\H_\text{loc}$-transformations of $\upphi$, implying that the dressed theory is $\G_{[\upphi]}$-(quasi-)invariant. 
In other words, the dressed formulation automatically implements \emph{physical frame~covariance}. 

Now, since the number of fields (d.o.f.) is limited, for all practical purposes (and realistic models), $\G_{[\upphi]}$ will be \emph{discrete}, reflecting the finite number of possible 
choices of fields (d.o.f.) that can serve as dressing fields.
Yet, for each dressing there will be a continuous $\G_{[\varnothing]}$-family. 
Since $\G_{[\varnothing]}$ is continuous, one may worry that an (internal) hole-type argument could be formulated for it. 
We do not regard this as a serious problem for either the DFM framework or the central thesis of this paper. First, it could simply be answered formally by another (internal) point-coincidence argument. More importantly, one would be hard-pressed to find any compelling reason to regard $\G_{[\varnothing]}$ as anything more than reflecting an unavoidable formal freedom encoding no information of physical or conceptual relevance.\footnote{In contrast to the original symmetry ($\Diff(M)\ltimes$)$\H_\text{loc}$ which, as  argued in earlier sections, encodes the relational structure of (gR)GFT \cite{JTF-Ravera2024c}.} 
Perhaps reinforcing this point, notice that each dressing gives a \emph{distinct} $\G_{[\varnothing]}$ action: e.g. for  
as $\C\left((\upphi_1)^u\right)=0$ and  $\C'\big((\upphi_2)^{u'}\big)=0$ above, one would get a $\G_{[\varnothing]}^1$-action for $u_1$-dressed fields, but a $\G_{[\varnothing]}^2$-action for $u_2$-dressed fields. 
In our view, it is clear that the mathematical, physical, and conceptual essence lies in the physical frame covariance implemented by the discrete group $\G_{[\upphi]}$, which 
cannot give rise to a hole-type argument.
\smallskip

Similarly to the internal case, given their general definition \eqref{dressing-field-diffeo}-\eqref{dressing-field-diffeo-field-dep},  two $\Diff(M)$-dressing fields are \emph{a priori} related by 
$\upsilon' = \upsilon \circ \zeta \rdefeq \upsilon^\zeta$, with $\zeta \in \mathcal{N}\defeq\{ \zeta \in \Diff(N)\, |\,  \zeta^\psi =\zeta \}$, and $\mathcal{N}$ is a group under composition. 
This gives an action on dressed fields $(\upphi^\upsilon)^\zeta \defeq (\upphi^\zeta)^{\upsilon\circ \zeta}$, with 
 special case $(\upphi^\upsilon)^\zeta:= (\upphi)^{\upsilon\circ \zeta}= \zeta^* (\upphi^\upsilon)$ if $\mathcal N$ leaves $\upphi$ invariant---in which case the dressed theory is $\mathcal{N}$-covariant, for the same formal reason that the bare theory is $\Diff(M)$-covariant. 
See \cite{Francois2023, JTF-Ravera2024gRGFT}.
All the above discussion applies \emph{mutatis mutandis}: 
formal ambiguities in the constructive procedure(s) to build dressing field(s) give rise to the continuous group(s) $\mathcal N^{(i)}_{[\varnothing]}$ (with $i=\{1, 2, \ldots\}$), while physical frame covariance of the dressed formulation is implemented by the discrete group $\mathcal N_{[\upphi]}$, for which no hole argument can arise.
Let us briefly consider some illustrative examples.
\medskip

One sometimes encounters the claim that QED admits an `infinity of dressings' built from the gauge potential~$\upphi=A$.
This is an instance of the formal ambiguity discussed above, namely $\U(1)_{[\varnothing]}$ arising from a  constraint $\C(A^u)=0$ solved for $u=u[A]$.
This, we notice, is often understood as `a dressing arising from a gauge fixing';\footnote{To the best of our knowledge, this view can be traced back to the authors of \cite{McMullan-Lavelle97}, who  also systematically introduced the notion and terminology of dressing, independently of \cite{Francois2014, GaugeInvCompFields}.} in this regard, we refer to the above discussion on the distinction between DFM and gauge fixing. 
See, in particular, \cite{Berghofer-Francois2024}, where the case of the `Lorenz gauge'-type constraint $\C(A^u)=\d^\mu(A_\mu^u)=0$ is discussed in detail. 

Working within $\CC$-scalar QED (or EM), with fields $\upphi=\{A, \vphi\}$, where $\vphi$ is a $\CC$-valued (charged) matter field, one may extract a $\U(1)$-dressing $u=u[\vphi]$ from the polar decomposition $\vphi=\rho e^{i\theta}\rdefeq \rho u[\vphi]$, where $\rho=|\vphi^u|$ is the modulus of $\vphi$. 
A possible ambiguity arises from the fact that one may insert  $1=\xi\-\xi$ in $\vphi=\rho \xi\-\xi u[\vphi]$ so that $u'[\vphi]=u[\phi]\xi$. 
Taking $\xi$ to be $U(1)$-valued, the ambiguity group is $\U(1)'_{[\varnothing]}$.
One may object that this does not respect the polar decomposition assumption (into a $\RR^+$-valued modulus and a phase), which is enforced by writing the constraint to be solved for $u$ as $\C(\vphi^u)=\vphi^u/|\vphi^u|-1=0$, for $\vphi^u\defeq u\-\vphi$---which is a `unitary gauge'-like condition (see the previous comment).
This forces $\xi=1$, meaning that there is no non-trivial ambiguity group here, and the dressing $u=u[\vphi]$ is thus unique. 
Now consider a model with two matter fields,  $\upphi=\{A, \vphi_1, \vphi_2\}$. One then obtains two unique dressings, $u=u[\vphi_1]$ and $u'=u'[\vphi_2]$. The internal reference frame covariance of the dressed formulations is therefore implemented by $\U(1)_{[\upphi]}=\{\xi[\upphi]=u\-u' \}$, shifting between  the frames, or `perspectives', of one matter field to the other.\footnote{We remark that this encompasses technically the gauge-invariant account of (scalar) electrodynamics of \cite{Wallace2014}, which is called  ``field monism" in \cite{Jacobs2022} and whose flavor of relationalism we sympathize with. 
Yet, \cite{Jacobs2022} rejects this approach on two grounds: 1) that it is not readily generalizable beyond the Abelian case, and 2) that it suffers from possible ambiguities which, in his view, are ``hardly better than" those associated with gauge symmetries. We think that, as discussed above, the DFM answers both criticisms.}

Consider also the example of tetrad formulations of GR, where $\upphi=\{A, e\}$,
 with $A$ the Lorentz connection and $e$ the tetrad field, both supporting $\H_\text{loc}=\SO(1,3)$-gauge transformations, which yet leave the theory $L_\text{GR}(A, e)$ invariant. 
 The
 $GL$-valued $\SO$-dressing field $u=u[e]={e^a}_\mu$ is obtained by solving   $\C(e^u)= e^u-dx^{\,\mu}=0$, with $e^u\defeq u\- e$.
 The $\SO$-invariant dressed fields $\Gamma\defeq A^u$ and $dx\defeq u\-e$ are the linear connection and the basis 1-form, respectively. 
 Yet, notice that baked in the constraint is a choice of \emph{coordinates}; another would have led to $u'[e]=u[e]\xi$ for $\xi: M \rarrow GL(4)$, or, in more usual terms, ${e^a}_\nu = {e^a}_\mu \, {\xi^\mu}_\nu$. 
 This means that the ambiguity group is $\GL_{[\varnothing]}$, which just encodes arbitrary coordinate changes, under which indeed dressed fields $\Gamma$ and $dx$ transform just like $A$ and $e$ transform under $\SO$, and under which the dressed theory (`metric' GR) remains invariant.
 Remember also that, as noted in footnote \ref{Stachel-active-diff}, per Stachel \cite{Stachel1986, EarmanNorton, Stachel2014} there is  \emph{no hole argument for mere coordinate changes} $\GL_{[\varnothing]}$, a.k.a. `passive' diffeomorphisms (see also \cite{JTF-Ravera2024c}). 

 Consider GR once again, now coupled to a collection of scalar fields (i.e. phenomenological matter fields, as in cosmology for example), so that $\upphi=\{g, \vphi\}$ with $\vphi=\vphi^i$ and $i=\{1, \ldots, 4\}$. The $\Diff(M)$-dressing field $\upsilon=\upsilon[\vphi]\defeq \vphi\-: N\subset \RR^4 \rarrow M$, with $N$ the value set (image) of $\vphi$, is immediately identified, or found, by solving the constraint
 $\C(\varphi^\upsilon)= \upsilon^* \varphi  -id_N=0$. The dressed metric, namely the relational variable  $g^\upsilon$, is the gravitational field as defined by the reference frame of the matter field, while $\vphi^\upsilon=\id_N$ is the matter field's representation of itself in its own rest frame. 
 This construction encompasses various forms of scalar coordinatization via ``material reference frames", e.g. à la DeWitt or Brown–Kucha\v{r} \cite{DeWitt1960,Kuchar1980, Rovelli1991, Brown-Marolf1996} (see also \cite{JTF-Ravera2025bdy}). 
 An ambiguity arises because the constraint is also satisfied by $\upsilon'[\vphi]=\upsilon[\vphi] \circ \zeta$, where $\zeta \in \N_{[\varnothing]} \subset \Diff(\RR^4)$. 
 Clearly, $\N_{[\varnothing]}$ represents arbitrary virtual coordinates that one is free to extrapolate from the existing physical reference frame given by $u[\vphi]$.\footnote{To form a mental picture that can help illustrate this, one may imagine a gravitational-wave interferometer as the physical reference frame supplying $\upsilon[\varphi]$ and $g^\upsilon$ the physically measured gravitational pulse. Then, $\N_{[\varnothing]}$ encodes all virtual coordinate systems one may anchor to, or extrapolate from, the physical interferometer.} 
 In a model with distinct matter fields $\upphi=\{g, \vphi_1, \varphi_2\}$, one thus has two candidate dressing fields $\upsilon[\vphi_1]$ and $\upsilon'[\vphi_2]$, with distinct (but isomorphic) ambiguity groups $\N^1_{[\varnothing]}$ and $\N^2_{[\varnothing]}$, while the physical frame covariance group is $\N_{[\upphi]}=\{\zeta[\upphi]\defeq \upsilon^{-1}\circ \upsilon'\}$, and allows to shift from the reference frame of one matter field to the other.\footnote{Such is the situation e.g. in the context of (relativistic) multifluid hydrodynamics 
 \cite{Andersson-Comer2007, Rezzolla-Zanotti2013}, where  fluid distributions are described by a set of scalar fields labelling fluid particles, known as ``Lagrangian coordinate fields" or ``comoving coordinates", while standard, unphysical coordinates on $M$ are referred to as ``Eulerian coordinates".}
 \smallskip

The freedom in the choice of dressing
does not undermine the picture developed in sections \ref{The case of internal gauge symmetries}-\ref{The case of general-relativistic gauge field theories} of an enriched physical spacetime for gRGFT represented by a dressed bundle space 
 $(P^{\,\u}, \omega^\u ,\alpha^\u )$,  locally  seen as 
 $(M^\upsilon, \upphi^{(\upsilon, u)} )$.
Its multiple, discrete realizations via  distinct dressing fields reflect  genuinely distinct physical perspectives, permuted via the discrete groups of 
 physical frame covariance $(\mathcal{N}_{[\upphi]},  \G_{[\upphi]})$. 
We shall soon revisit this topic more extensively  \cite{Berghofer-et-al2027-C}.

\section{Conclusion}
\label{Conclusion}

In this paper, we have outlined a framework for interpreting general-relativistic Gauge Field Theory (gRGFT) in which its covariance group of local transformations, $\Diff(M)\ltimes\H_\text{loc}$, is seen---via the articulation of the generalized hole and point-coincidence arguments---to tacitly encode its fundamental relational structure. Yet, the ontological picture presented here does not follow the canvas set by the traditional substantivalist versus relationalist dichotomy. 

Rather, it strongly suggests a relationalist realist view of a principal bundle as representing the manifold of physical spatio-temporal and internal events, 
which supervene on---or supervene as a ``structural quality" of---the fundamental physical fields of the theory. 
The d.o.f. of these fields are coextensive with the mutually co-defining  network of their (invariant) relations (section \ref{The hole argument} and \ref{The generalized hole argument}).
This nicely aligns with a ``moderate" form of ontic structural realism  regarding the fundamental fields of gRGFT; a form of OSR which, in contrast to \cite{French}, is not eliminativist regarding relata (``objects") and does not favor a \mbox{primary} ontology of relations (see \cite{sep-structural-realism,Berghofer2018}) but regards relata and relations as inseparable and on equal-footing ontologically. 
\enlargethispage{1\baselineskip}
This form of structural realism traces back to \cite{Eddington1929}, as noted by \cite{French2003} and \cite{Rickles2008}, and has been further developed specifically in the context of fiber bundles under the label of ``dynamic structural realism" by \cite{Stachel2014}.

Where we go beyond Stachel and others is by using the dressing field method (DFM), showing how dressed fields, bundles, and spaces are immune to any form of (generalized) hole argument. 
We indeed contend that the manifestly invariant and manifestly relational reformulation of gRGFTs via the DFM 
(section \ref{Dressing Field Method: Dressed fields and spaces}) is as close a technical realization of (moderate) ontic structural realist views as currently possible within the standard framework of differential bundle geometry (section \ref{Fiber bundles and basics of gauge theory}).  
The dressed fields \eqref{dressed-fields}, \eqref{diffeo-dressed-fields}, \eqref{dressed-fields-gRGFT} and the dressed regions \eqref{dressed-regions}, \eqref{dressed-regions-gRGFT} of the manifold of events $M^\upsilon$  and of the enriched manifold of events $P^{\,\u}\!\rarrow\! M^\upsilon$,  formally implement the generalized point-coincidence argument, and are by construction immune to the generalized hole argument.
This, then, (re)opens the door for fiber bundle substantivalist views.

The ``bundle-realism" elaborated here has several interesting interpretive implications,
for example regarding the electroweak model (and the notion of spontaneous symmetry breaking), and the Aharonov-Bohm (AB) effect. 
Indeed, the view elaborated above rests crucially on the insight of the generalized point-coincidence argument, 
and is then a completely local (in the field-theoretic sense) understanding of the physics of gRGFT. 
The AB effect is thus interpretable in a gauge-invariant local way, as a phenomenon of differential parallel transport \emph{in the physical bundle space} $P^{\,\u}$ resulting from the local interaction of the physical matter field $\phi^\u$ and the physical electromagnetic potential $\omega^\u$---even though the effect is seen to be ``displayed"   effectively on the (quotient) space $M^\upsilon$ between the local representative fields $\vphi^{(\upsilon, u)}$ and $A^{(\upsilon, u)}$, which are ``shadows" of the true play. 
The AB effect is thus not essentially different, and no more mysterious, than comparable parallel transport effects in GR inducing gravitational time dilation---such as the Shapiro effect, the gravitationally induced phase shift in photon, or even the Langevin twins effect---and resulting from the local interaction between  physical matter and interaction fields ($\vphi^{(\upsilon, u)}$/$A^{(\upsilon, u)}$) and the physical gravitational potential~($g^\upsilon$). 
Both types of effects can be understood in the same terms, and neither requires a form of ``holism".\footnote{The term is used by, e.g., \cite{Nounou2003, Lyre2004, Healey2007} to denote a form of spatio-temporal non-locality or non-separability, and should not be confused with the better-known quantum variant often used in the context of entanglement.}
{We will expand on these terse remarks in a separate contribution \cite{Berghofer-et-al2027-B} discussing the AB effect through the lens of the generalized hole and point-coincidence argument and, expanding on  \cite{JTF-Ravera2024c}, framing it within the DFM approach. }
Relatedly, while \cite{Healey2007} took holonomies (i.e. integrals of potentials $A$ along curves/loops in $M$) as a favorite candidate for the fundamental ontology of (gR)GFT, it is clear that on the view proposed here, holonomies are derived quantities, and---like the physical bundle space, its regions, points and curves---supervene  on the fundamental relational ontology of physical (dressed) fields.

\section*{Acknowledgment}  

P.B. acknowledges that his research was funded by the Austrian Science Fund (FWF), grant \mbox{[P 36542]} and grant \mbox{[AST1252624]}.
J.F. is supported by the Austrian Science Fund (FWF), grant \mbox{[P 36542]}, and by the Czech Science Foundation (GAČR), grant GA24-10887S.
L.R. is supported by the research grant PNRR Young Researchers, MSCA Seal of Excellence (SoE), CUP E13C24003600006, ID SOE2024$\_$0000103, project GrIFOS.
We thank two careful referees for their constructive remarks that pushed us to give precisions at various points,  making the manuscript clearer.

{
\normalsize 
 \bibliography{references}
}

\end{document}